\documentclass[11pt]{article}

\usepackage[utf8]{inputenc}
\usepackage[title]{appendix}
\usepackage{bm}
\usepackage{tikz}
\usetikzlibrary{arrows.meta, positioning}
\usetikzlibrary{fit}
\usetikzlibrary{calc}
\usetikzlibrary{backgrounds}
\usepackage{amsmath}
\usepackage{amssymb}
\usepackage{listings}
\usepackage{graphicx}
\usepackage[font=footnotesize]{subcaption}
\usepackage{colortbl}
\PassOptionsToPackage{table,xcdraw,x11names,rgb}{xcolor}
\usepackage{xcolor}
\usepackage{rotating}
\usepackage{multicol}
\usepackage{multirow}
\usepackage{enumitem}
\usepackage{placeins}
\usepackage{anyfontsize}
\usepackage[most]{tcolorbox}
\usepackage{adjustbox}
\usepackage{geometry}
\usepackage{icomma}
\usepackage{longtable}
\usepackage[numbers,super,sort&compress]{natbib}
\usepackage{titlesec}
\usepackage{caption}
\usepackage{booktabs}
\usepackage{url}

\titleformat{\paragraph}
{\normalfont\normalsize\bfseries}{\theparagraph}{1em}{}
\titlespacing*{\paragraph}
{0pt}{3.25ex plus 1ex minus .2ex}{1.5ex plus .2ex}
\usepackage[colorinlistoftodos]{todonotes}
\usepackage{enumitem}

\usepackage[pdftex,pagebackref=false]{hyperref} 
\hypersetup{
    plainpages=false,       
    unicode=false,          
    pdftoolbar=true,        
    pdfmenubar=true,        
    pdffitwindow=false,     
    pdfstartview={FitH},    
    pdftitle={A Tutorial on Conducting Mediation Analysis with Exposure Mixtures},    
    pdfauthor={Yiran Wang, Yi-Ting Lin, Sean McGrath, John D Meeker, Sung Kyun Park, Joshua L Warren, Bhramar Mukherjee},    
    pdfnewwindow=true,      
    colorlinks=true,        
    linkcolor=blue,         
    citecolor=blue,        
    filecolor=magenta,      
    urlcolor=cyan           
}

\tcbuselibrary{listings}

\newtcblisting{rcodebox}{
  listing only,
  colback=gray!5!white, colframe=black!85,
  listing options={
    language=R,
    basicstyle=\ttfamily\footnotesize,
    keywordstyle=\color{black},
    commentstyle=\color{gray},
    stringstyle=\color{black},
    breaklines=true,
    columns=flexible
  },
  boxrule=1pt,
  arc=1mm,
  left=10pt,
  right=10pt,
  top=0pt,
  bottom=0pt,
}

\DeclareMathOperator*{\argmin}{argmin}
\newcommand{\E}{\text{E}}
\newcommand{\xs}{x^\ast}

\newcommand{\bbeta}{\bm{\beta}}
\newcommand{\balpha}{\bm{\alpha}}
\newcommand{\iid}{\overset{iid}{\sim}}
\newcommand{\ind}{\mathbin{\perp\mspace{-10mu}\perp}}
\newcommand{\given}{\!\mid\!}
\newcommand{\X}{\bm{X}}
\newcommand{\x}{\bm{x}}
\newcommand{\xxs}{\bm{x}^\ast}
\newcommand{\C}{\bm{C}}
\newcommand{\bc}{\bm{c}}
\newcommand{\bS}{\bm{S}}
\newcommand{\bss}{\bm{s}^\ast}
\newcommand{\bs}{\bm{s}}
\newcommand{\Z}{\bm{Z}}
\newcommand{\z}{\bm{z}}

\usepackage{authblk}

\title{A Tutorial on Conducting Mediation Analysis with Exposure Mixtures}

\author[1]{Yiran Wang$^{\dagger\ast}$}
\author[2]{Yi-Ting Lin$^\ast$}
\author[1]{Sean McGrath}
\author[3]{John D. Meeker}
\author[3,4]{Sung Kyun Park}
\author[1]{Joshua L. Warren}
\author[1,5]{Bhramar Mukherjee}

\affil[1]{Department of Biostatistics, Yale School of Public Health, New Haven, Connecticut, USA}
\affil[2]{Department of Biostatistics, School of Public Health, University of Michigan, Ann Arbor, Michigan, USA}
\affil[3]{Department of Environmental Health Sciences, School of Public Health, University of Michigan, Ann Arbor, Michigan, USA}
\affil[4]{Department of Epidemiology, School of Public Health, University of Michigan, Ann Arbor, Michigan, USA}
\affil[5]{Department of Chronic Disease Epidemiology, Yale School of Public Health, New Haven, Connecticut, USA}

\date{}

\begin{document}
\maketitle

\abstract{Causal mediation analysis is a powerful tool in environmental health research, allowing researchers to uncover the pathways through which exposures influence health outcomes. While traditional mediation methods have been widely applied to individual exposures, real-world scenarios often involve complex mixtures. Such mixtures introduce unique methodological challenges, including multicollinearity, sparsity of active exposures, and potential nonlinear and interactive effects. This paper provides an overview of several commonly used approaches for mediation analysis under exposure mixture settings with clear strategies and code for implementation. The methods include: single exposure mediation analysis (SE-MA), principal component-based mediation analysis (PC-MA), environmental risk score-based mediation analysis (ERS-MA), and Bayesian kernel machine regression causal mediation analysis (BKMR-CMA). While SE-MA serves as a baseline that analyzes each exposure individually, the other methods are designed to address the correlation and complexity inherent in exposure mixtures. For each method, we aim to clarify the target estimand and the assumptions that each method is making to render a causal interpretation of the estimates obtained. We conduct a simulation study to systematically evaluate the operating characteristics of these four methods to estimate global indirect effects and to identify individual exposures contributing to the global mediation under varying sample sizes, effect sizes, and exposure-mediator-outcome structures. We also illustrate their real-world applicability by examining data from the PROTECT birth cohort, specifically analyzing the relationship between prenatal exposure to phthalate mixtures and neonatal head circumference Z-score, with leukotriene E4 as a mediator. This example offers practical guidance for conducting mediation analysis in complex environmental contexts.}


\renewcommand\thefootnote{}
\footnotetext[1]{$^\dagger$Corresponding author: Yiran Wang (\href{mailto:yiran.wang.yw995@yale.edu}{yiran.wang.yw995@yale.edu}).}
\footnotetext[2]{$^\ast$Y.W. and Y.-T.~L. contributed equally to this work.}

\renewcommand\thefootnote{\fnsymbol{footnote}}
\setcounter{footnote}{1}

\clearpage

\section{Introduction}\label{intro}

Causal mediation analysis is a powerful statistical framework for investigating the mechanisms, or pathways, through which an exposure may affect an outcome.~\cite{vanderweele2016mediation} Under appropriate assumptions, this framework decomposes the total effects of an exposure into two components: the direct effect, which represents the exposure's impact on the outcome when mediators are held constant, and the indirect effect, which operates through one or more mediators.

Over the last two decades, as genetics moved from single-gene to multi-gene and genome-wide models, environmental health has likewise shifted from one-at-a-time analyses to studying exposure mixtures with an eye toward the exposome. Compared with genomics, mixture analyses in environmental settings bring distinct challenges: exposures are lower in dimension but continuous, often measured with error, correlated over time, and subject to latency; the exposure–outcome relation may be nonlinear and involve interactions.~\cite{dominici2010protecting,toccalino2012chemical,wang2018associations,henn2012associations,claushenn2014chemical}

A variety of methods now exist to characterize the joint effect of mixtures,~\cite{hamra2018environmental} but this characterization is only a first step. Once an overall mixture effect is established, one may want to investigate mechanisms by decomposing the mixture’s effect into direct and indirect components with respect to a mediator---for example, biomarkers along specific metabolic pathways---to generate biological insight and inform potential intervention targets.~\cite{aung2020application} Methodological work on mediation analysis with mixtures is growing, yet remains heterogeneous in estimands and assumptions, which can complicate interpretation and cross-study comparison. This tutorial responds by providing a structured, practice-oriented overview.

Against this background, we review four approaches spanning three broad classes for mediation analysis with multiple environmental toxicants measured simultaneously. These methods vary in their assumptions and modeling strategies, and not all of them support formal causal interpretation. Some aim to estimate causal mediation effects under explicit identification conditions, while others provide descriptive or associational insights into mediation pathways. The four approaches of interest include:

\begin{itemize}[label=$\circ$]
    \item \textbf{Single exposure mediation analysis (SE-MA)}:\\
    Applies standard mediation models to each exposure separately. While simple and widely used, this approach ignores the correlation among exposures, and, as a result, yields biased estimates of individual indirect effects due to lack of adjustment for co-exposures, and cannot recover the global indirect effect of the mixture unless exposures are uncorrelated or perfectly collinear.
    \item \textbf{Mediation analysis based on exposure summaries}:
    \begin{itemize}[label=$\cdot$]
        \item \textbf{Principal component-based mediation analysis (PC-MA)}:~\cite{zitko1994principal,reid2009use,ma2012principal,abdi2010principal} Reduces correlated exposures into uncorrelated components representing major sources of joint variation within the mixture, and carries out mediation analysis with one or more leading principal components.
        \item \textbf{Environmental risk score-based mediation analysis (ERS-MA)}:~\cite{wang2018associations,park2014environmentala}Constructs a weighted summary score of selected (or all) exposures to represent overall exposure burden, and conducts mediation analysis using the ERS as the single exposure.
    \end{itemize}
    \item \textbf{Bayesian kernel machine regression causal mediation analysis (BKMR-CMA)}:~\cite{bobb2015bayesiana,devick2022bayesiana}\\
    A flexible, semiparametric framework that models nonlinear and interactive mixture effects nonparametrically via a kernel while retaining parametric components, directly estimates causal mediation effects through Markov chain Monte Carlo (MCMC) sampling, and supports variable selection (e.g., component-wise and hierarchical) to identify influential exposures or exposure groups.
\end{itemize}

This tutorial paper has four primary objectives. First, it introduces a set of commonly used methods for conducting mediation analysis with exposure mixtures. Second, it discusses the key identification assumptions and underlying causal structures of each approach. Third, it provides practical guidance on how to implement these methods, including step-by-step instructions and code blocks. Fourth, it evaluates the performance of these methods through simulation studies, and highlights their relative strengths and limitations to guide users in selecting appropriate tools based on study goals and data characteristics. The aim of this paper is to help researchers understand the trade-offs involved in considering mediation analysis with mixtures. This paper does not introduce any new methodology; rather, it serves as a resource for applied researchers seeking to navigate and apply existing techniques in environmental health studies.

We illustrate the application of these methods using data from the Puerto Rico Testsite for Exploring Contamination Threats (PROTECT) project. This prospective cohort study investigates the impact of maternal environmental chemical exposures on birth outcomes. The dataset includes 40 chemical exposures across four major classes: phthalates, phenols and parabens, polycyclic aromatic hydrocarbons, and metals. The primary outcome of interest for this study is the head circumference Z-score, a standardized measure of fetal head growth at birth linked to neurodevelopmental outcomes. Potential mediating pathways are assessed through biomarkers, specifically metabolites from the cytochrome P450 and lipoxygenase pathways, which reflect biological responses to chemical exposures.~\cite{aung2021cross} Specifically, we focus our illustration on 11 phthalates within PROTECT as exposures with a single mediator (LTE4) and head circumference Z-score as the outcome. Details are described in Section~\ref{protect}.

The structure of the paper is as follows. Section~\ref{method} introduces the methods for causal mediation analysis in the presence of exposure mixtures, including SE-MA, PC-MA, ERS-MA, and BKMR-CMA. Section~\ref{simulation} evaluates the performance of each method using simulated datasets, focusing on their ability to estimate mediation effects and identify active exposures under varying sample sizes and effect strengths. Section~\ref{protect} presents a case study using the PROTECT dataset to illustrate real-world applications. Section~\ref{discussion} discusses key findings, methodological trade-offs, and practical guidance for selecting among mediation methods in complex exposure settings. To promote transparency and reproducibility, detailed implementation code for all methods is available at~\url{https://github.com/ysph-dsde/Mixture_Mediation_Tutorial}.

\section{Methods}\label{method}

\subsection{Mediation Analysis}\label{mediation}

Mediation analysis is a statistical framework used to understand the mechanisms through which an exposure influences an outcome. Specifically, it decomposes the total effect of an exposure on the outcome into two key components: the direct effect, which captures the portion of the exposure’s effect that bypasses intermediate variables (mediators), and the indirect effect, which operates through mediators that lie on the causal pathway between the exposure and the outcome.~\cite{robins1992identifiability,pearl2001direct,vanderweele2009conceptual,vanderweele2010odds,imai2010generala,valeri2013mediation} 

In this paper, we focus on the setting with multiple exposures, a single continuous mediator, and a continuous outcome. Let the observed data consist of $n$ independent observations of the outcome $Y$, the mediator $M$, a vector of $p$ exposures $\X$, and a vector of $s$ confounders $\C$ that may influence any part of the exposure-mediator-outcome pathway. Throughout this paper, we use bold symbols such as $\X$ and $\C$ to represent vector-valued variables at the individual level. For example, each individual may have multiple exposures or confounders, which are denoted using bold letters. Scalar quantities, such as the outcome $Y$, are written without bold. When bold symbols are accompanied by explicit dimensions, such as $\X_{(n\times p)}$, they refer to full data matrices across all $n$ individuals.

\begin{figure}[!htb]
    \centering
    \begin{subfigure}[t]{0.9\textwidth}
        \centering
        \begin{tikzpicture}[
            node distance = 0.6cm and 0.8cm,
            every node/.style = {align=center, font=\scriptsize},
            arrow/.style = {-{Stealth}, thick}
        ]   
        \node[draw=blue, rounded corners] (X) at (0,0) {Exposures \\ 
        $\displaystyle \X_{(n \times p)} =\qquad\qquad\qquad$ \\$ \left\{(\X)_{1_{(n \times p_1)}}, \dots, (\X)_{g_{(n \times p_g)}}\right\}$};
        \node[draw=blue, rounded corners] (M) [right=of X] {Mediator \\ 
        $\displaystyle M_{(n \times 1)}$};
        \node[draw=blue, rounded corners] (Y) [right=of M] {Outcome \\ 
        $\displaystyle Y_{(n \times 1)}$};
        \node[draw=blue, rounded corners] (C)  at ($(X)!0.5!(Y)+(0,2.8)$) {Confounders \\
        $\displaystyle \C_{(n \times s)}$};        
        
        \draw [arrow] (X) -- (M);
        \draw [arrow] (M) -- (Y);
        \draw [arrow] (X) to[bend right] (Y);
        \draw [arrow] (C) -- (X);
        \draw [arrow] (C) -- (M);
        \draw [arrow] (C) -- (Y);
        
        \end{tikzpicture}
        \caption{General directed acyclic graph (DAG) illustrating the relationships between exposures ($\X$), mediator ($M$), confounders ($\C$), and outcome ($Y$). Exposures are grouped into $g$ mixture components, where $(\X)_j$ represents the exposures in the $j$-th component.}
        \label{fig:mediation}
    \end{subfigure}
    \vspace{1em}
    \begin{subfigure}[t]{0.9\textwidth}
        \centering
        \begin{tikzpicture}[
            node distance = 0.6cm and 0.8cm,
            every node/.style = {align=center, font=\scriptsize},
            arrow/.style = {-{Stealth}, thick}
        ]    
        \node[draw=brown, rounded corners] (X) at (0,0) {Phthalates \\ $\displaystyle
        \begin{aligned}
        \X_{(175\times 11)} &=
        \\[-2pt]
        &\hspace{-1em}\bigl\{\, (\X)_{HMW_{(175\times 8)}},\\
        &\hspace{-1em}\quad (\X)_{LMW_{(175\times 3)}} \,\bigr\}
        \end{aligned}$};
        \node[draw=brown, rounded corners] (M) [right=of X, xshift=-0.2cm] {Leukotriene E4\\ 
        $\displaystyle M_{(175 \times 1)}$};
        \node[draw=brown, rounded corners] (Y) [right=of M, xshift=-0.2cm] {Head\\ Circumference\\ Z-score \\ 
        $\displaystyle Y_{(175 \times 1)}$};
        \node[draw=brown, rounded corners] (C) [above=of M, yshift=1cm, xshift=-1cm] {Maternal Covariates \\
        $\displaystyle \C_{(175 \times 3)}$};

        \draw [arrow] (X) -- (M);
        \draw [arrow] (M) -- (Y);
        \draw [arrow] (X) to[bend right] (Y);
        \draw [arrow] (C) -- (X);
        \draw [arrow] (C) -- (M);
        \draw [arrow] (C) -- (Y);
        
        \end{tikzpicture}
        \caption{DAG corresponding to the PROTECT dataset example, where phthalate exposures are denoted by $\X=\{(\X)_{HMW}, (\X)_{LMW}\}$, grouped into high and low molecular weight categories based on their chemical properties. Leukotriene E4 (LTE4) is the mediator $M$, head circumference Z-score is the outcome $Y$, and maternal characteristics (age, education, BMI) are confounders $\C$. This grouping is used solely for illustrative purposes in this figure.}
        \label{fig:mediation_protect}
    \end{subfigure}

    \caption{Directed acyclic graphs (DAGs) illustrating the mediation framework for exposure mixtures. The upper panel represents the general framework, while the lower panel applies this framework to the PROTECT dataset.}
    \label{fig:mediation_combined}
\end{figure}
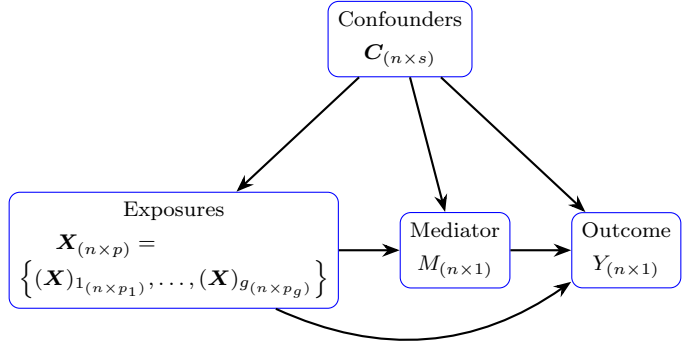
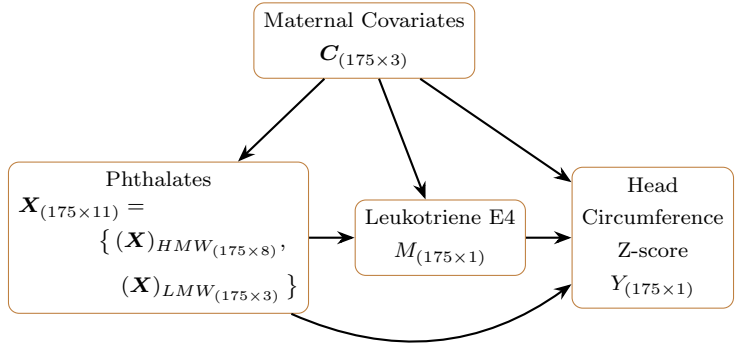

We illustrate this framework using data from the PROTECT study, which examines the effects of environmental chemical exposures on birth outcomes. Here, the exposures $\X$ consist of 11 correlated phthalate metabolites measured in 175 pregnant participants. To facilitate illustration in the DAG (Figure~\ref{fig:mediation_protect}), these metabolites are grouped into high molecular weight (HMW) and low molecular weight (LMW) phthalates based on their chemical structure and common sources. The HMW group includes MCPP, MCOP, MCNP, MBzP, MEHP, MEHHP, MEOHP, and MECPP, while the LMW group includes MEP, MBP, and MiBP. Note that this grouping is used only for illustration and is not applied in the data analysis. The outcome $Y$ is the head circumference Z-score at birth, a marker of fetal development. The mediator $M$ is Leukotriene E4 (LTE4), a biomarker from the lipoxygenase pathway. Maternal characteristics, including age, education, and pre-pregnancy BMI, are treated as confounders $\C$. Additional details on the PROTECT dataset are provided in Section~\ref{protect}.

\subsubsection{Traditional Mediation Analysis with Multiple Exposures}

Under this setting, practitioners often begin by analyzing each exposure individually. If the exposures are uncorrelated, the single-exposure approach provides a straightforward strategy for identifying potential mediation pathways. However, in many real-world applications, particularly in environmental health, exposures occur as mixtures, representing groups of highly correlated variables that may share common sources or act through similar biological mechanisms. Ignoring these dependencies can lead to biased inferences, especially when exposures exhibit multicollinearity or jointly influence the mediator or outcome. 

In the context of exposure mixtures, the $p$ exposures in $\X$ are often grouped into $g$ non-overlapping components, denoted $(\X)_1,\ldots,(\X)_g$, where each group reflects shared characteristics (e.g., chemical class or source) or clustered correlation structure. Let $p_i$ be the number of exposures in group $(\X)_i$, so that the total number of exposures is given by $p=\sum_{i=1}^gp_i$. Figure~\ref{fig:mediation} provides a general directed acyclic graph (DAG) representing the mediation structure under an exposure mixture setting. 

\subsubsection{Causal Mediation Analysis with Multiple Exposures}

To formalize the causal effects in such settings, we adopt the counterfactual (or potential outcomes) framework, following the work of Robins and Greenland,~\cite{robins1992identifiability} and Pearl.~\cite{pearl2001direct} Let $Y(\x)$ denote the potential outcome if the exposures $\X$ were set to $\x$, and $M(\x)$ denote the potential mediator value under the same exposure level. If exposures $\X$ were set to $\x$ and mediator $M$ were set to $m$, the potential outcome is denoted by $Y(\x,m)$. 

Using this notation, we define population-average causal effects corresponding to different pathways. When comparing an exposure level $\x$ to a reference level $\xxs$, the (average) natural direct effect (NDE) is defined as 
$$\E[Y(\x, M(\xxs)) - Y(\xxs, M(\xxs))],$$
which quantifies how the outcome would change if the exposures shift from $\xxs$ to $\x$, while the mediator is held at the level it would have naturally taken under the reference exposure level $\xxs$. The (average) natural indirect effect (NIE) is defined as
$$\E[Y(\x, M(\x)) - Y(\x, M(\xxs))],$$
which captures the change in outcome when the mediator changes from the level under $\xxs$ to that under $\x$, while holding the exposures constant at $\x$. These definitions correspond to the pure NDE and the total NIE in some literature. Under the assumptions detailed below, the total effect (TE) decomposes as the sum of NIE and NDE: 
$$\text{NDE + NIE} = \E[Y(\x) - Y(\xxs)],$$
representing the overall impact of changing the exposure mixture from $\xxs$ to $\x$. In addition to the natural effects, another estimand of interest is the (average) controlled direct effect (CDE), defined as 
$$\E[Y(\x, m) - Y(\xxs, m)],$$
which assesses the effect of changing the exposure mixture from $\xxs$ to $\x$ while fixing the mediator at a specific value $m$. Unlike the natural effects, the CDE corresponds to hypothetical interventions that set the mediator to a fixed level, and is often used to evaluate direct effects under policy-relevant or clinical scenarios.. 

\subsubsection{Identification Assumptions}

The validity of estimation for causal effects relies on several key assumptions for identification. Throughout this section, identification is stated conditional on measured covariates $\C$. If population-average effects are of interest, they can be obtained by marginalizing over the distribution of $\C$. First, the consistency assumption links potential outcomes to observed variables: if an individual receives exposure $\X = \x$ and mediator $M = m$, then their observed outcome $Y$ equals $Y(\x,m)$, and their observed mediator $M$ equals $M(\x)$.~\cite{vanderweele2009conceptual} A related composition assumption states that $Y(\x,M(\x))=Y(\x)$ for all $\x$.~\cite{ding2024first} Second, the positivity assumption ensures adequate data support across covariate strata, requiring that all exposure and mediator levels occur with positive probability: $P(\X=\x\given\C=\bc)>0$ for all $\x,\bc$ and $P(M=m\given \C=\bc, \X=\x)>0$ for all $\x, \bc, m$.~\cite{lange2017applied} Third, the stable unit treatment value assumption (SUTVA) requires no interference between individuals and no hidden versions of treatment or mediator.~\cite{rubin1990formal}

For the identification of CDEs, it is sufficient to assume no unmeasured confounding of the exposure-outcome and mediator-outcome relationships, conditional on measured covariates $\C$.~\cite{vanderweele2015explanation} The CDE can then be identified as
\begin{align*}
    \text{CDE}(\x,\xxs\given\C)&=\E[Y(\x, m) - Y(\xxs, m)\given\C]\\
    &=\E[Y\given\X=\x,M=m,\C] \\
    &\phantom{=}- \E[Y\given\X=\xxs,M=m,\C].
\end{align*}

To identify NDEs and NIEs, two additional assumptions are needed. The first is no unmeasured confounding of the exposure-mediator relationship.~\cite{devick2022bayesiana,pearl2001direct,vanderweele2009conceptual} Together with the conditions required for CDEs, these three no-unmeasured-confounding conditions are commonly referred to as sequential ignorability,~\cite{imai2010identification} which can be expressed as: (1) $Y(\x, m) \ind \X \given \C$, (2) $M(\x) \ind \X \given \C$, and (3) $Y(\x, m) \ind M \given \X, \C$. The second additional assumption is the cross-world independence assumption: $Y(\x,m)\ind M(\xxs)\given \C$ for all $\x$, $\xxs$, and $m$. This assumption implies that the exposures do not affect any mediator-outcome confounders. However, it is untestable from observed data and often viewed as strong, since $Y(\x, m)$ and $M(\xxs)$ correspond to counterfactuals under different exposure levels and can never be jointly observed when $\x \neq\xxs$.~\cite{ding2024first}

Under these assumptions, causal mediation effects can be identified as
\begin{align*}
    \text{NDE}(\x,\xxs\given\C)&=\E[Y(\x, M(\xxs)) - Y(\xxs, M(\xxs))\given\C]\\
    &=\int\{\E[Y\given\X=\x,M=m,\C]\\
    &\phantom{=}-\E[Y\given\X=\xxs,M=m,\C]\}\\
    &\phantom{=}\times P(M=m\given\X=\xxs,\C)dm,\\
    \text{NIE}(\x,\xxs\given\C)&=\E[Y(\x, M(\x)) - Y(\x, M(\xxs))\given\C]\\
    &=\int\E[Y\given\X=\x,M=m,\C]\times\\
    &\phantom{=}\{P(M=m\given\X=\x,\C)\\
    &\phantom{=}- P(M=m\given\X=\xxs,\C)\}dm.
\end{align*}

While these assumptions form the basis for identifying causal mediation effects, specific estimation methods may impose additional assumptions, such as parametric model forms or restrictions on the data-generating process. It is also important to note that the definition of ``exposure'' varies across the methods considered in this paper. For instance, in PC-MA, the exposures correspond to principal component scores; in ERS-MA, the exposure is the environmental risk score constructed from the mixture; and in BKMR-CMA, the entire mixture is modeled nonparametrically. The assumptions described above apply to the effective exposure representation specific to each method. 

\subsubsection{Regression-based Estimation with Product and Difference Methods} \label{prod_diff}

Several regression-based strategies have been developed to estimate the causal effects defined above. We focus here on two classical approaches: the product method and the difference method.~\cite{vanderweele2016mediation}

The difference method estimates the indirect effect by first fitting two separate regression models to estimate the total effect and the natural direct effect:
\begin{align*}
    \E[Y\given\X,\C] &= \phi_0 + \X\phi_x + \C\phi_c,\\
    \E[Y\given\X,M,\C]&= \beta_0 + \X\beta_x + M\beta_m + \C\beta_c.
\end{align*}
The first model captures the total effect of the exposures on the outcome, conditional on the confounders. When comparing two exposure levels $\x$ and $\xxs$, this effect is given by 
$$\E[Y\given \X=\x,\C] - \E[Y\given \X=\xxs,\C] = (\x-\xxs)\phi_x.$$
The second model includes the mediator, and estimates the natural direct effect as
\begin{align*}
    \E[Y\given\X=\x, M=M(\xxs),\C]-&\E[Y\given\X=\xxs, M=M(\xxs),\C]\\
    &=(\x - \xxs)\beta_x.
\end{align*}

The natural indirect effect is then obtained by subtraction: $(\x - \xxs)(\phi_x - \beta_x)$. This difference can also be interpreted as the change in the exposure effect after adjusting for the mediator.

The product method estimates mediation effects by modeling the effect of the exposures on the mediator, and then the joint effects of exposures and the mediator on the outcome. This approach involves the following mediator model and outcome model:
\begin{align*}
    \E[M\given \X, \C] &= \alpha_0 + \X\alpha_x + \C\balpha_c,\\
    \E[Y\given \X, M, \C] &= \beta_0 + \X\beta_x + M\beta_m + \C\bbeta_c.
\end{align*}
Under this framework, the total effect of shifting exposures from $\xxs$ to $\x$ is given by $(\x-\xxs)(\beta_x+\alpha_x\beta_m)$, which decomposes into the direct effect $(\x-\xxs)\beta_x$ and the indirect effect $(\x-\xxs)\alpha_x\beta_m$. The indirect effect arises as the product of the effect of exposures on the mediator and the effect of the mediator on the outcome, motivating the term ``product method.'' This decomposition holds when the models are correctly specified, the exposure-mediator interaction is absent, and the relevant identification assumptions are satisfied. In our setting, because both the mediator and the outcome are continuous and we fit linear models without an exposure-mediator interaction, the product and difference methods are algebraically equivalent and yield identical estimates of the natural direct and indirect effects. However, this is not the case when linearity or the no-interaction assumption is violated, or when using models with nonlinear link functions, such as logistic regression for a binary outcomes.~\cite{vanderweele2016mediation}

In the remainder of the tutorial, we default to the product method for estimation in SE-MA, PC-MA, and ERS-MA (with linear models and no exposure–mediator interaction). BKMR-CMA is a semiparametric approach: it models the exposure mixture nonparametrically via a kernel function while retaining parametric components. Its mediation analysis relies on the mediator and outcome models introduced above, together with a regression for the total effect of the exposures, paralleling the models introduced in the product and difference methods.

To make the code examples interoperable across methods, we organize the analysis dataset as follows:
\begin{tcolorbox}
\begin{verbatim}
data <- data.frame(
      X1 = exposure_1,
      ...,
      Xp = exposure_p,
      M = mediator,     
      Y = outcome,      
      C1 = confounder_1,
      ...,
      Cs = confounder_s
)
\end{verbatim}
\end{tcolorbox}

Here, each exposure and confounder appears as a separate column in the data frame, consistent with the input requirements for the methods implemented below.

\subsection{Single Exposure Mediation Analysis}\label{sema}

As described in the previous section, one practical approach for mediation analysis in exposure mixtures is to evaluate each exposure individually. We refer to this method as single exposure mediation analysis (SE-MA). In this section, we adopt the product method introduced in Section~\ref{prod_diff}, primarily following the work of Baron and Kenny and its extensions.~\cite{vanderweele2015explanation,baron1986moderator} 

In practice, researchers may follow one of two strategies: (1) fitting the mediator and outcome models without adjusting for other exposures in the mixture, or (2) including the remaining exposures as additional covariates in both models. While both strategies adopt the same two-model regression structure, they differ substantially in terms of causal interpretation and potential sources of bias.

\begin{figure}[!htb]
    \centering
    \begin{subfigure}[t]{\linewidth}
    \centering
    \begin{tikzpicture}[
        node distance = 1.2cm and 1.8cm,
        every node/.style = {align=center, font=\scriptsize},
        arrow/.style = {-{Stealth}, thick}
    ]    
    \node[draw=blue, rounded corners] (X) at (0,0) {Exposure \\ 
    $X_{j \; (n\times1)}$};
    \node[draw=blue, rounded corners] (M) [right=of X] {Mediator \\ 
    $M_{(n\times1)}$};
    \node[draw=blue, rounded corners] (Y) [right=of M] {Outcome \\ 
    $Y_{(n\times1)}$};
    \node[draw=blue, rounded corners] (C) [above=of M, yshift=1cm] {Confounders \\
    $\C_{(n\times s)}$};
    \node[draw=blue, rounded corners] (U) [above=of M, left=of C] {Unadjusted Co-Exposures \\
    $\X_{-j \; (n\times(p{-}1))}$};
    
    \draw [red,dashed] (U) -- (X);
    \draw [arrow,red] (U) -- (M);
    \draw [arrow,red] (U) -- (Y);
    \draw [arrow] (X) -- (M);
    \draw [arrow] (M) -- (Y);
    \draw [arrow] (X) to[bend right] (Y);
    \draw [arrow] (C) -- (X);
    \draw [arrow] (C) -- (M);
    \draw [arrow] (C) -- (Y);
    
    \end{tikzpicture}
    \caption{Directed acyclic graph illustrating mediation analysis of a single exposure $X_j$ without adjustment for co-exposures. The remaining exposures $\X_{-j}$ are treated as variables that may be associated with $X_j$ and may confound the mediator-outcome relationship. Red arrows indicate unadjusted pathways that can introduce bias in the estimated causal effects. The dashed line represents a non-causal association between co-exposures and the exposure of interest.}
    \label{fig:ind_dag}
\end{subfigure}
\hfill
\begin{subfigure}[t]{\linewidth}
    \centering
    \begin{tikzpicture}[
        node distance = 1.2cm and 1.8cm,
        every node/.style = {align=center, font=\scriptsize},
        arrow/.style = {-{Stealth}, thick}
    ]    
    \node[draw=brown, rounded corners] (X) at (0,0) {Exposure \\ 
    $X_{j \; (n\times1)}$};
    \node[draw=brown, rounded corners] (M) [right=of X] {Mediator \\ 
    $M_{(n\times1)}$};
    \node[draw=brown, rounded corners] (Y) [right=of M] {Outcome \\ 
    $Y_{(n\times1)}$};
    \node[draw=brown, rounded corners] (C) [above=of M, yshift=1cm] {Confounders \\
    $\C_{(n\times s)}$};
    \node[draw=brown, rounded corners] (U) [above=of M, left=of C] {Adjusted Co-Exposures \\
    $\X_{-j \; (n\times(p{-}1))}$};
    
    \draw [dashed] (U) -- (X);
    \draw [arrow] (U) -- (M);
    \draw [arrow] (U) -- (Y);
    \draw [arrow] (X) -- (M);
    \draw [arrow] (M) -- (Y);
    \draw [arrow] (X) to[bend right] (Y);
    \draw [arrow] (C) -- (X);
    \draw [arrow] (C) -- (M);
    \draw [arrow] (C) -- (Y); 
    
    \end{tikzpicture}
    \caption{Directed acyclic graph illustrating mediation analysis of a single exposure $X_j$ with adjustment for co-exposures $\X_{-j}$. All relevant paths are accounted for, enabling valid estimation of the causal mediation effect of $X_j$ under standard identification assumptions. The dashed line represents a non-causal association between co-exposures and the exposure of interest.}
    \label{fig:ind_co_dag}
\end{subfigure}
\caption{Directed acyclic graphs comparing single exposure mediation analysis (SE-MA) with and without adjustment for co-exposures. }
\end{figure}
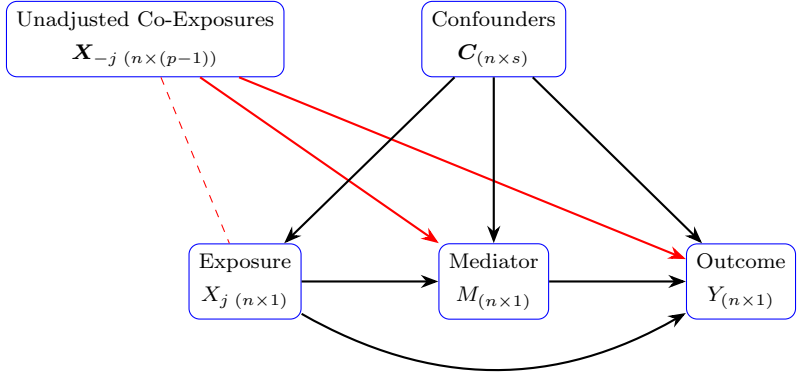
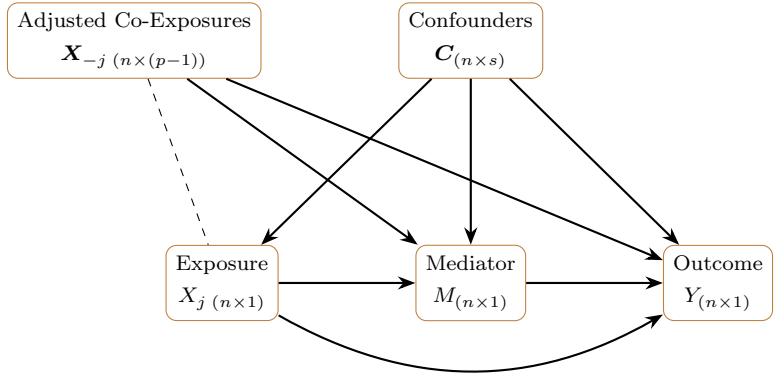

When co-exposures are excluded, the exposure of interest may exhibit associations with other mixture components that themselves influence the mediator or outcome, as shown in Figure~\ref{fig:ind_dag}. In this setting, although co-exposures are measured, they are not adjusted for, and thus act as omitted confounders for the mediator-outcome relationship, violating the assumption that all confounding variables are appropriately controlled. As a result, the estimated mediation effects for single exposures lack causal interpretability. Furthermore, even aggregating these biased estimates across exposures does not recover a valid global effect of the mixture, since each estimate is derived under a misspecified and incompatible causal structure. This approach is, therefore, fundamentally flawed for estimating either individual or global causal effects in the presence of co-exposures.

Suppose $X_j$ is the exposure of interest from the mixture $\X$. When both the mediator and the outcome are continuous, SE-MA uses the same mediator and outcome models introduced above, now specified for a single exposure $X_j$:
\begin{equation*}
    \E(M\given X_j, \C) = \alpha_0 + X_j\alpha_x + \C\balpha_c,
\end{equation*}
where the coefficient $\alpha_x$ captures the effect of the exposure on the mediator, adjusting for confounders. And the outcome model is
\begin{equation*}
    \E(Y\given M, X_j, \C) = \beta_0 + X_j\beta_x + M\beta_m + \C\bbeta_c,
\end{equation*}
where $\beta_x$ represents the direct effect of the exposure on the outcome, and $\beta_m$ measures the effect of the mediator on the outcome, controlling for the exposure and confounders.

Based on these two models, NDE, NIE, and TE of exposure $X_j$ can be estimated as
\begin{align}
    \widehat{\text{NDE}}(x_j,\xs_j\given\C) &= (x_j-\xs_j)\hat{\beta}_x,\label{eq:nde}\\
    \widehat{\text{NIE}}(x_j,\xs_j\given\C) &= (x_j-\xs_j)\hat{\beta}_m\hat{\alpha}_x,\label{eq:nie}\\
    \widehat{\text{TE}}(x_j,\xs_j\given\C) &= (x_j-\xs_j)(\hat{\beta}_x + \hat{\beta}_m\hat{\alpha}_x),\label{eq:te}
\end{align}
where $\hat{\beta}_x$, $\hat{\beta}_m$ and $\hat{\alpha}_x$ are the estimated coefficients from the outcome and mediator models. The quantities $x_j$ and $\xs_j$ denote the comparative and reference level of the exposure $X_j$, respectively. We assume there is no exposure-mediator interaction, although it can be relaxed by including an interaction term in the outcome model.~\cite{vanderweele2016mediation}

When co-exposures are included as covariates, as shown in Figure~\ref{fig:ind_co_dag}, the analysis adjusts for potential confounding due to other exposures in the mixture that are associated with the exposure of interest and influence the mediator, the outcome, or both. This adjustment enables valid causal interpretation of the effect for the exposure of interest, under standard identification assumptions. Specifically, this structure assumes that one exposure varies while all others in the mixture are held fixed, treating co-exposures as control variables. This setting allows for the identification of exposure-specific causal effects, but does not support causal interpretation on the effect of the mixture as a whole. Because one mediation model is fit per exposure, testing exposure-specific NIEs requires multiple-testing correction; common choices include family-wise error rate control and false discovery rate procedures.~\cite{dunn1961multiple,holm1979simple,benjamini1995controlling} In the simulations in Section~\ref{simulation}, we use the Benjamini–Hochberg procedure to control the FDR when flagging potentially active exposures.~\cite{benjamini1995controlling} 

Although the same model could also be used to estimate the global effect of the entire mixture, the estimation of global causal effects requires the structure shown in Figure~\ref{fig:mediation}, in which all mixture components are considered jointly as exposures rather than covariates. Furthermore, because exposures in mixtures are often highly correlated, adjusting for co-exposures may introduce multicollinearity, even under a correctly specified model. In such cases, while the causal interpretation remains valid in theory, practical inference may be unstable. 

To formalize this setting, we again consider a single exposure of interest $X_j$ from the mixture $\X$, with the remaining exposures denoted by $\X_{-j}$. The mediator model is specified as
$$\E(M\given X_j,\X_{-j},\C)=\alpha_0+X_j\alpha_x+(\X_{-j}, \C)\balpha_{c},$$
and the outcome model is
$$\E(Y\given M,X_j,\X_{-j},\C)=\beta_0+X_j\beta_x+M\beta_m+(\X_{-j}, \C)\bbeta_{c}.$$ The NDE, NIE, and TE of $X_j$ are then estimated using the same expressions in Equations~\eqref{eq:nde},~\eqref{eq:nie}, and~\eqref{eq:te}.

While SE-MA offers a simple and interpretable approach, it is conceptually limited in the context of exposure mixtures. When co-exposures are excluded, the models are misspecified and yield biased estimates that lack causal interpretation. When co-exposures are adjusted for as confounders, the estimated effects of the single exposure can be interpreted causally under certain identification assumptions. Alternatively, if all exposures are modeled jointly as components of the mixture rather than treated as confounders, the resulting model can be used to estimate valid global effects. Although the statistical models may take similar forms in both scenarios, they correspond to distinct causal contrasts: the former estimates the effects of varying one exposure while holding others fixed, whereas the latter captures the global effect of changing the entire mixture. In addition, the high correlation among exposures can introduce multicollinearity, which undermines the precision and reliability of inference. Therefore, while SE-MA serves as a useful baseline, it should be applied with caution in mixture settings, where the causal role of co-exposures must be clearly specified to ensure valid interpretation.

\subsubsection{Code for SE-MA}

In this paper, we implement SE-MA using the \texttt{cmest} function from the R package \texttt{CMAverse}.~\cite{shi2021cmaverse} The following code illustrates how to estimate mediation effects when both the mediator and outcome models are linear and no exposure-mediator interaction is included:
\begin{rcodebox}
# Perform single exposure mediation analysis (no co-exposure scenario)
est <- CMAverse::cmest(
  data = data,                 # Data frame containing the variables
  model = "rb",                # Model type ("rb" for regression-based mediation analysis)
  outcome = "Y",               # Name of the outcome variable
  exposure = "Xj",             # Name of the exposure variable
  mediator = "M",              # Name of the mediator variable
  basec = c("C1", ...,"Cs"),   # Confounders excluding those affected by exposure
  EMint = FALSE,               # No exposure-mediator interaction in the outcome model
  mreg = list("linear"),       # Linear mediator model
  yreg = "linear",             # Linear outcome model
  estimation = "paramfunc",    # Use closed-form parameter function estimation
  inference = "delta"          # Use delta method for inference
)
\end{rcodebox}

In the above code, \texttt{data} is the input dataset containing all required variables. The arguments \texttt{outcome}, \texttt{exposure}, \texttt{mediator}, and \texttt{basec} specify the names of the corresponding variables in the dataset. Here, \texttt{Xj} should be replaced by the name of the specific exposure variable of interest from the mixture. The \texttt{EMint} argument controls whether exposure-mediator interaction is included in the outcome model, and is set to \texttt{FALSE} for this example. The \texttt{estimation} argument specifies the estimation approach: \texttt{"paramfunc"} implements closed-form parameter function estimation, while \texttt{"imputation"} uses direct counterfactual imputation. The \texttt{inference} argument controls uncertainty quantification, with \texttt{"bootstrap"} (nonparametric bootstrap resampling) and \texttt{"delta"} (delta method) being the two main options. If \texttt{"bootstrap"} is used, the number of resamples must be specified via \texttt{nboots}. In all subsequent code examples we use \texttt{estimation="paramfunc"} and \texttt{inference="delta"}, which align with the analytic derivations presented earlier.

Besides the regression-based estimator used above (\texttt{model = "rb"}), the \texttt{cmest} function also implements several common alternatives that can be invoked by changing the \texttt{model} argument: \texttt{"wb"} (weighting-based estimator);~\cite{vanderweele2014mediation} \texttt{"iorw"} (inverse odds–ratio weighting);~\cite{tchetgen2013inverse} \texttt{"ne"} (natural effect models);~\cite{vansteelandt2012imputation} \texttt{"msm"} (marginal structural models);~\cite{vanderweele2017mediation} and \texttt{"gformula"} (parametric g-formula).~\cite{robins1986new}

To adjust for co-exposures, other exposure variables can be added to the \texttt{basec} argument. For example, when \texttt{X1} is the exposure of interest:
\begin{rcodebox}
# Perform single exposure mediation analysis (with co-exposure scenario)
est <- CMAverse::cmest(
  data = data,                 # Data frame containing the variables
  model = "rb",                # Model type ("rb" for regression-based mediation analysis)
  outcome = "Y",               # Name of the outcome variable
  exposure = "X1",             # Name of the exposure variable
  mediator = "M",              # Name of the mediator variable
  basec = c("C1",...,"Cs","X2",...,"Xp"),  # Confounders and co-exposures
  EMint = FALSE,               # No exposure-mediator interaction in the outcome model
  mreg = list("linear"),       # Linear mediator model
\end{rcodebox}
\begin{rcodebox}
  yreg = "linear",             # Linear outcome model
  estimation = "paramfunc",    # Use closed-form parameter function estimation
  inference = "delta"          # Use delta method for inference
)
\end{rcodebox}

\subsection{Principal Component-Based Mediation Analysis}\label{pca}

Principal component analysis (PCA) is a dimension reduction technique that transforms a set of potentially correlated exposures into a smaller set of uncorrelated components, known as principal components (PCs).~\cite{gibson2019overview} In this paper, we use unsupervised PCA, meaning that the transformation is performed without reference to the outcome or mediator. This transformation captures the dominant patterns of variability in the exposures while mitigating multicollinearity.~\cite{abdi2010principal, jolliffe2016principal} By reducing the dimensionality of the exposure data, PCA preserves the most informative structures, allowing the analysis to focus on key components that explain the majority of the variance.

The PCA process begins by standardizing the exposure mixture $\X$, ensuring that all exposures have a mean of zero and a variance of one. Standardization is essential to prevent exposures with larger variances from dominating the analysis. Next, PCA is performed by either eigenvalue decomposition of the covariance matrix or singular value decomposition of the standardized exposure matrix. The resulting eigenvectors, also known as PC loadings, define the new orthogonal directions in the data space along which the variance is maximized. Each eigenvalue quantifies the variance explained by its corresponding PC. Typically, the first few PCs capture the largest portion of the total variance, allowing researchers to reduce the dimensionality of the data while retaining most of the information. PCA can be implemented in R using the \texttt{prcomp} function, with standardization and centering options to ensure consistent scaling across exposures.~\cite{R} A scree plot can be used to visualize the proportion of variance explained by each principal component, aiding the selection of the number of PCs to retain. We provide example code for conducting PCA and generating a scree plot in Section~\ref{pca_code}.

After selecting the PCs, the original exposure variables are projected onto these components to generate PC scores for each observation. These scores, denoted $\bS$, serve as substitutes for the original exposures $\X$ and can be incorporated into the \texttt{cmest} function to estimate causal mediation effects. The mediator and outcome models with PC scores are given by:
\begin{align*}
    \E[M\given \bS, \C] &= \alpha_0^{\text{PCA}} + \bS\balpha_s^{\text{PCA}} + \C\balpha_c^{\text{PCA}},\\
    \E[Y\given M, \bS, \C] &= \beta_0^{\text{PCA}}+\bS\bbeta_s^{\text{PCA}}+M\beta_m^{\text{PCA}}+\C\bbeta_c^{\text{PCA}}.
\end{align*}
Figure~\ref{fig:pca_dag} provides a causal diagram corresponding to this modeling approach, where the PC scores $\bS=f(\X)$ act as surrogates for the original exposures. Once the models are fitted, the causal mediation effects can be determined for the PC scores. Specifically, the NDE, NIE, and TE are expressed as:
\begin{align*}
    \widehat{\text{NDE}}_{\text{PCA}}(\bs,\bss\given\C) &= (\bs-\bss)\hat{\bbeta}_x^{\text{PCA}},\\
    \widehat{\text{NIE}}_{\text{PCA}}(\bs,\bss\given\C) &= (\bs-\bss)\hat{\balpha}_x^{\text{PCA}}\hat{\beta}_m^{\text{PCA}},\\
    \widehat{\text{TE}}_{\text{PCA}}(\bs,\bss\given\C) &= (\bs-\bss)(\hat{\bbeta}_x^{\text{PCA}}+\hat{\balpha}_x^{\text{PCA}}\hat{\beta}_m^{\text{PCA}}),
\end{align*}
where $\bs$ and $\bss$ are the comparative and reference levels of the PC scores, respectively. The dimensions of the resulting coefficients depend on the number of PCs retained and the number of observations ($n$). If $l$ PCs are retained, then $\bS$ has dimension $n\times l$. Consequently, $\hat{\balpha}_x^{\text{PCA}}$ and $\hat{\bbeta}_x^{\text{PCA}}$ have a dimension of $l \times 1$ and $\hat{\beta}_m^{\text{PCA}}$ is a scalar.

An important consideration when applying PCA to exposure mixtures is the presence of exposure groupings or clustering, often suggested by block-like correlation patterns in the exposure matrix. In such settings, the leading principal components frequently align with these groupings, which can simplify interpretation. Using the resulting PC scores as surrogates for the original exposures allows for the estimation of the global mediation effect of the mixture in a lower-dimensional space. However, this approach relies on the implicit assumption that the retained PCs adequately summarize the information in the original exposures, such that the mediator and outcome are conditionally independent of $\X$ given the PC scores and confounders ($Y, M \ind \X \given \bS, \C$).~\cite{boss2023shrinkage} 

In practice, only the first few PCs are typically retained, which can lead to violations of these assumptions if important exposure information is omitted. Moreover, although causal effects can be formally defined for the PC scores, it is important to recognize that each score represents a linear combination of multiple original exposures. As a result, intervening on a single PC score while holding others fixed may not correspond to a plausible or physically meaningful intervention, complicating the causal interpretation at the individual component level. Causal interpretation is generally more appropriate when considering the joint effect across the set of retained components. Furthermore, because causal effects are estimated with respect to derived components rather than the original exposure variables, the results are difficult to map back directly to individual exposures. These limitations highlight the trade-off between dimensionality reduction and causal interpretability when using PCA in mediation analysis of exposure mixtures.

\subsubsection{Code for PC-MA}\label{pca_code}

In R, PCA can be performed using the \texttt{prcomp} function, and the resulting PC scores can be used as exposures in mediation analysis. The following code illustrates how to perform PC-MA in R:

\begin{rcodebox}
# Step 1: Perform unsupervised PCA on the standardized exposure matrix
res_pca <- prcomp(X, center = TRUE, scale. = TRUE)

# Step 2: Visualize explained variance to decide how many PCs to retain
fviz_eig(res_pca)

# Step 3: Extract PC scores that explain 80% of the total variance
explained_variance <- summary(res_pca)$importance[3, ]  # Cumulative variance
id_80var <- which.max(explained_variance >= 0.8)        # Index of first PC exceeding 80%
PC_scores <- res_pca$x[, 1:id_80var]                    # Subset of retained PC scores

# Step 4: Append PC scores to the dataset
colnames(PC_scores) <- paste0("PC", 1:id_80var)
data <- cbind(data, PC_scores)

# Step 5: Perform mediation analysis using the first PC as the exposure
est <- CMAverse::cmest(
  data = data,                     # Data frame with all relevant variables
  model = "rb",                    # Regression-based mediation analysis
  outcome = "Y",                   # Outcome variable
  exposure = "PC1",                # Exposure of interest (first PC score)
\end{rcodebox}
\begin{rcodebox}
  mediator = "M",                  # Mediator variable
  basec = c("C1", ..., "Cs",       # Baseline confounders
            paste0("PC", 2:id_80var)),  # Remaining PCs as covariates
  EMint = FALSE,                   # No exposure-mediator interaction
  mreg = list("linear"),           # Linear mediator model
  yreg = "linear",                 # Linear outcome model
  estimation = "paramfunc",    # Use closed-form parameter function estimation
  inference = "delta"          # Use delta method for inference
)
\end{rcodebox}

This code starts by performing PCA on the exposure matrix \texttt{X} using the \texttt{prcomp} function. The argument \texttt{center = TRUE} centers each exposure to have mean zero, and \texttt{scale. = TRUE} scales them to unit variance—both are essential for performing PCA on data with variables on different scales. 

A scree plot can be generated using \texttt{fviz\_eig} from the \texttt{factoextra} package to help determine the number of PCs to retain.~\cite{factoextra} Common selection criteria include explaining a cumulative variance threshold (e.g., 70\%-90\%) or retaining PCs with eigenvalues greater than one (Kaiser criterion).~\cite{kaiser1960application} In this example, we select the smallest number of components that together explain at least 80\% of the total variance. The corresponding PC scores are then extracted from \texttt{res\_pca\$x}, renamed to \texttt{PC1}, \texttt{PC2}, ..., and appended to the original dataset.

To estimate the effect of a specific component, such as the first PC in this example, we use the \texttt{cmest} function from the \texttt{CMAverse} package, treating the remaining PCs and baseline confounders as covariates. This approach yields estimates of the natural direct, indirect, and total effects for the selected PC, under the assumption that the retained PCs adequately capture the variation in the original exposures. 

If a global mediation effect under a joint shift in the retained PC scores is of interest, the corresponding effect can be obtained by summing the PC-specific effects for that shift. This provides a reasonable approximation when most exposure information is captured by the selected components, although variation in omitted PCs is not accounted for.

\subsection{Environmental Risk Score-Based Mediation Analysis}\label{ers}

The environmental risk score (ERS) is a scalar metric designed to summarize the joint effects of multiple exposures in an environmental mixture on a health outcome.~\cite{wang2018associations,park2014environmentala,park2017constructiona} Unlike unsupervised dimension reduction techniques such as PCA, ERS uses the outcome $Y$ to guide the estimation, making it a supervised method. By integrating multiple correlated exposures into a single score, the ERS simplifies mediation analysis while preserving critical information about the exposure-outcome relationship.

A variety of methods can be used to construct the ERS, depending on the complexity of the exposure-outcome relationship. Nonparametric and ensemble-based approaches such as Bayesian additive regression trees~\cite{chipman2010bart}, Bayesian kernel machine regression~\cite{bobb2015bayesiana}, and super learner~\cite{van2007super} offer flexible alternatives capable of capturing nonlinearities and interactions among exposures.~\cite{park2017constructiona} While more flexible methods may be preferred in practice, we illustrate the ERS construction using the elastic net, which provides a transparent and interpretable approach suitable for demonstration purposes. The elastic net combines LASSO ($L^1$) and ridge regression ($L^2$) penalties to perform variable selection and shrinkage.~\cite{zou2005regularization} To avoid overly sparse scores dominated by a few exposures, we require the inclusion of at least three exposures in the final model. Penalty parameters are selected via five-fold cross-validation to minimize squared error loss.

To reduce overfitting and ensure valid downstream inference, we adopt a simple strategy in this paper by randomly splitting the data into two equally sized subsets: a training set and an analysis set. The elastic net model is fit on the training set to estimate exposure coefficients, which are then used to construct ERS in the analysis set. Let $i = 1, \ldots, n$ index subjects, and let $j=1,\ldots,p$ index the exposure variables. For each subject $i$, let $\Z_i=(X_{i1},\ldots,X_{ip},C_{i1},\ldots,C_{is})$ denote the full vector of predictors, including $p$ exposures and $s$ confounders. The elastic net estimates the coefficients by solving the following optimization problem:
\begin{equation*}
    \hat{\bbeta} = \argmin_{\bbeta}\left[\sum_{i=1}^n(Y_i-\Z_i\bbeta)^2+\lambda_1\sum_{j=1}^p\lvert\beta_j\rvert+\lambda_2\sum_{j=1}^p\beta_j^2\right],
\end{equation*}
where $\lambda_1$ and $\lambda_2$ are the penalties for $L^1$ and $L^2$ regularization, respectively. The penalties are applied only to the exposure coefficients $\beta_1,\ldots,\beta_p$, while the coefficients for the confounders are left unpenalized by assigning them zero penalty factors.

Depending on the modeling goal, the predictor vector $\Z_i$ can include different sets of variables. In this paper, we consider two versions of $\Z_i$ for constructing the ERS: one that includes only the main effects of the exposures, and another that incorporates squared terms and pairwise interactions to capture potential nonlinearities and synergistic effects. In both cases, all predictors are standardized prior to model fitting to ensure comparability across variables.

After fitting the elastic net on the training set, we use the estimated coefficients to construct the ERS for each subject in the analysis set. The ERS is computed as a linear combination of selected exposures, extended to include squared terms and pairwise interactions if such higher-order terms are specified in model training. The general form of the ERS for subject $i$ is given by:
\begin{equation*}
    \text{ERS}_i=\sum_{j=1}^p\hat{\beta}_jX_{ij} +  \sum_{k=1}^{p}\sum_{l=1}^p\hat{\beta}_{kl}X_{ik}X_{il},
\end{equation*}
where $X_{ij}$ denotes the $j$-th exposure for subject $i$, $\hat{\beta}_j$ is the estimated coefficient for the main effect of the $j$-th exposure, and $\hat{\beta}_{kl}$ is the estimated coefficient for the product term $X_{ik}X_{il}$, representing either a squared term ($k=l$) or a pairwise interaction ($k\neq l$). If only main effects are considered, the second term is omitted. If fewer than three exposures are selected at the optimal $\lambda_1$, we iteratively adjust $\lambda_1$ to ensure that at least three exposures are included in the final model. Note that due to penalization, many coefficients may be shrunk exactly to zero, leading to a sparse representation of the exposure-response relationship.

The resulting ERS replaces the original exposures in the mediation analysis framework, serving as a scalar summary of the exposure mixture. Subsequently, the \texttt{cmest} function can be employed once again to perform mediation analysis using the ERS as the exposure. The mediator and outcome models incorporating the ERS are specified as:
\begin{align*}
    \E[M\given \text{ERS}, \C] &= \alpha_0^{\text{ERS}} + \text{ERS}\alpha_{ers}^{\text{ERS}} + \C\balpha_c^{\text{ERS}},\\
    \E[Y\given M, \text{ERS}, \C] &= \beta_0^{\text{ERS}}+\text{ERS}\beta_{ers}^{\text{ERS}}+M\beta_m^{\text{ERS}}+\C\bbeta_c^{\text{ERS}}.
\end{align*}
The natural direct effect (NDE), natural indirect effect (NIE), and total effect (TE) of the ERS are then calculated as:
\begin{align*}
    \widehat{\text{NDE}}_{\text{ERS}}(\text{ERS},\text{ERS}^\ast\given\C) &= (\text{ERS}-\text{ERS}^\ast)\hat{\beta}_x^{\text{ERS}},\\
    \widehat{\text{NIE}}_{\text{ERS}}(\text{ERS},\text{ERS}^\ast\given\C) &= (\text{ERS}-\text{ERS}^\ast)\hat{\alpha}_x^{\text{ERS}}\hat{\beta}_m^{\text{ERS}},\\
    \widehat{\text{TE}}_{\text{ERS}}(\text{ERS},\text{ERS}^\ast\given\C) &= (\text{ERS}-\text{ERS}^\ast)(\hat{\beta}_x^{\text{ERS}}+\hat{\alpha}_x^{\text{ERS}}\hat{\beta}_m^{\text{ERS}}),
\end{align*}
where ERS and $\text{ERS}^\ast$ are the comparative and reference levels of the ERS score, respectively, and all coefficients involved in the effect estimates are scalars.

\begin{figure}[!hbt]
    \centering
    \begin{tikzpicture}[
        node distance = 0.5cm and 0.7cm,
        every node/.style = {align=center},
        arrow/.style = {-{Stealth}, thick}
    ]    
    \node[draw=brown, rounded corners] (E) at (0,0) {ERS \\ 
    $ERS=\hat{f}_C(\X_A)=\X_A\hat{\bbeta}_{\X|\C}$};
    \node[draw=brown, rounded corners] (M) [right=of E] {Mediator \\ 
    $M_A$};
    \node[draw=brown, rounded corners] (YA) [right=of M] {Outcome \\ 
    $Y_A$};
    \node[draw=brown, rounded corners] (CA) [above=of M, yshift=1cm] {Confounders \\
    $\C_A$};
    \node[draw=blue, rounded corners] (T) [left=1.8cm of E] (T) {$(Y, \X, \C)_{\text{Train}}$};

    \draw [arrow] (E) -- (M);
    \draw [arrow] (M) -- (YA);
    \draw [arrow] (E) to[bend right] (YA);
    \draw [arrow] (CA) -- (E);
    \draw [arrow] (CA) -- (M);
    \draw [arrow] (CA) -- (YA);
    \draw [dashed] (T) -- node[midway, above] {\(\hat{f}_C\)} (E);
    
    \begin{pgfonlayer}{background}
        \path (T.south west)+(-0.4,-0.5) node (a) {};
        \path (T.north east)+(0.4,0.5) node (b) {};
        \path[draw=black, thick, dashed, rounded corners]
            (a) rectangle (b);
        \node[font=\bfseries] at ($(a)!0.5!(b)+(0,1.2)$) {Training Set};
    
        \path (E.south west)+(-0.6,-1.2) node (c) {};
        \path (YA.north east)+(0.6,4.5) node (d) {};
        \path[draw=black, thick, dashed, rounded corners]
            (c) rectangle (d);
        \node[font=\bfseries] at ($(c)!0.5!(d)+(0,3.8)$) {Analysis Set};
    \end{pgfonlayer}
    
    \end{tikzpicture}
    \caption{Directed acyclic graph illustrating the use of environmental risk score (ERS) in mediation analysis of exposure mixtures. In the training set $(Y, X, C)_{\text{Train}}$, a supervised prediction model is fit to approximate $\E[Y\given\X,\C]$, yielding an estimated function $\hat{f}_C$. This function maps the exposures in the analysis set $X_A$ to a scalar ERS, computed as \( \hat{f}_C(\X_A)\). When the model includes only main effects, this reduces to a linear combination \( \hat{\bbeta}_{\X|\C} \); when nonlinear terms are also included, \( \hat{f}_C(\X_A)\) is computed based on transformed features of the exposures. The resulting ERS replaces the original exposures and enters as the exposure variable in mediation model, linking it to the mediator $M_A$ and outcome $Y_A$, with adjustment for confounders $C_A$. The dashed line from the training set to ERS represents the estimation of $\hat{f}_C$, rather than a direct causal relationship.}
    \label{fig:ers_dag}
\end{figure}
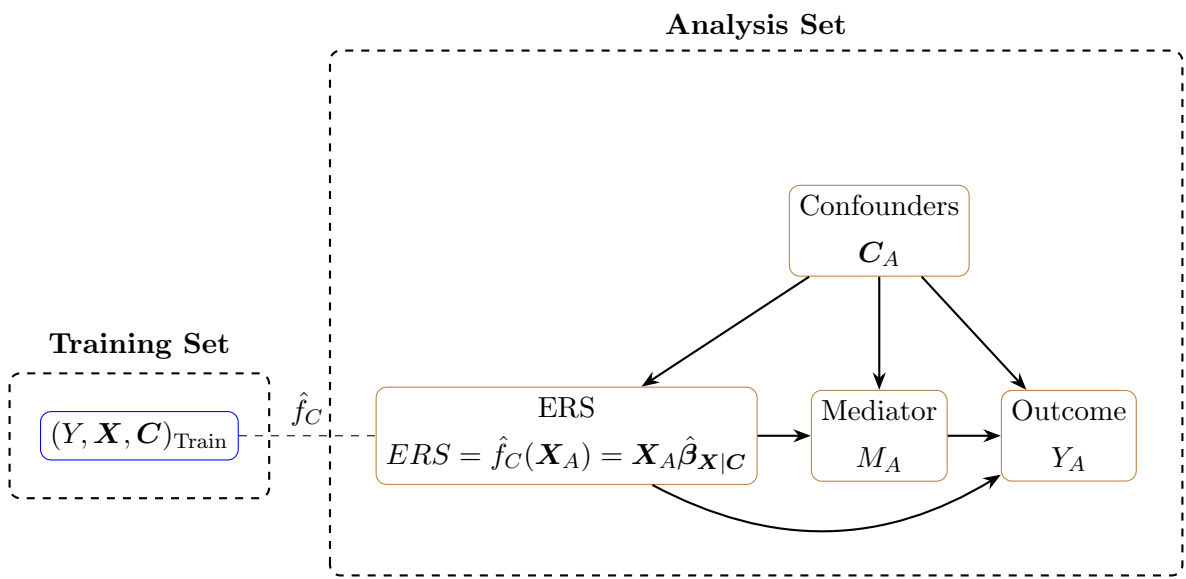

To summarize the procedure, Figure~\ref{fig:ers_dag} provides a schematic overview of the ERS-based mediation analysis framework. In the training set, we estimate a prediction function $\hat{f}_C$ that approximates $\E[Y\given\X,\C]$, using a supervised learning approach. In this paper, $\hat{f}_C$ is obtained via elastic net regularized regression, though other methods may also be applied. The estimated coefficients, denoted $\hat{\bbeta}_{\X|\C}$, are then applied to the exposures in the analysis set to compute the ERS. When only main effects are included, the ERS can be written as $\X_A\hat{\bbeta}_{\X|\C}$; when higher-order terms are included, the ERS is computed using a transformed set of predictors, but still represents a weighted sum of exposure features learned from the training data. The resulting ERS replaces the original exposures in downstream mediation models, summarizing the joint contribution of the exposure mixture through a single scalar.

The ERS serves as a global summary of the exposure mixture, enabling mediation analysis to estimate the overall direct and indirect effects of the combined exposures. While this approach provides a practical way to handle complex mixtures, it does not support estimation of individual mediation effects for specific exposures and instead focuses on the cumulative impact of the mixture. In addition, the validity of the analysis depends on the assumption that the ERS sufficiently captures all relevant information contained in the original exposures. This assumption is conceptually similar to the sufficiency requirement in PC-MA. 

Unlike PCA, the ERS is a supervised risk index tailored to the outcome of interest. A higher ERS indicates a higher risk of the outcome based on the combined effects of chemical mixtures, but it does not necessarily reflect higher levels of exposure to those mixtures. Its value also depends on the specific outcome used for model training. When external data are unavailable, estimating the exposure coefficients requires splitting the available dataset into a training set and an analysis set, which may reduce statistical power. As a one-dimensional score optimized for predicting the outcome, the ERS simplifies modeling but limits interpretability with respect to single exposure contributions and causal pathways.

\subsubsection{Code for ERS-MA}
To construct the ERS, we implement elastic net regularization with five-fold cross-validation in R using the \texttt{gcdnet} package.~\cite{gcdnet} The procedure involves splitting the data into a training and an analysis set, fitting an elastic net model in the training set with confounder adjustment, and applying the estimated coefficients to the analysis set to compute the ERS. Below is an illustrative code snippet demonstrating the key steps of this procedure with only main effects (no interactions or squared terms):
\begin{rcodebox}
# Load required library
library(gcdnet)

# Split data into training and analysis sets
n_train <- floor(nrow(data) / 2)
train_ids <- sample(seq_len(nrow(data)), n_train)
analysis_ids <- setdiff(seq_len(nrow(data)), train_ids)

X_train <- data[train_ids, paste0("X", 1:p)]
C_train <- data[train_ids, paste0("C", 1:s)]
Y_train <- data[train_ids, "Y"]

# Combine exposures and confounders from the training set into a single predictor matrix
Z_train <- cbind(X_train, C_train)

# Define penalty factors for L1 and L2 regularization: Exposures are penalized (1), confounders are unpenalized (0)
pf <- c(rep(1, ncol(X_train)), rep(0, ncol(C_train)))  # penalty factors

# Tune lambda2 and lambda1 using cross-validation
lambda2_values <- exp(seq(log(1e-4), log(1e2), length.out = 100)) # Define a range of lambda2 values for L2 regularization
cv_lambda2 <- sapply(lambda2_values, function(lambda2) {
    cv.gcdnet(
        x = Z_train,        # Predictor matrix (exposures + confounders)
        y = Y_train,        # Outcome variable
        lambda2 = lambda2,  # L2 regularization strength
        method = "ls",      # Use least-squares loss function
        nfolds = 5,         # Perform five-fold cross-validation
        pf = pf,            # Penalty factors for L1 regularization
        pf2 = pf            # Penalty factors for L2 regularization
    )$cvm                   # Extract cross-validated mean squared error
})
\end{rcodebox}
\begin{rcodebox}
optimal_lambda2 <- lambda2_values[which.min(cv_lambda2)]

# Select optimal lambda1 for the chosen lambda2
cv_result <- cv.gcdnet(
    x = Z_train,                # Predictor matrix
    y = Y_train,                # Outcome variable
    lambda2 = optimal_lambda2,  # Optimal L2 regularization strength
    method = "ls",              # Least-squares loss function
    nfolds = 5,                 # Five-fold cross-validation
    pf = pf,                    # Penalty factors for L1 regularization
    pf2 = pf                    # Penalty factors for L2 regularization
)
optimal_lambda1 <- cv_result$lambda.min

# Fit final elastic net model
best_model <- gcdnet(
    x = Z_train,               # Predictor matrix
    y = Y_train,               # Outcome variable
    lambda = optimal_lambda1,  # Optimal L1 regularization strength
    lambda2 = optimal_lambda2, # Optimal L2 regularization strength
    pf = pf,                   # Penalty factors for L1 regularization
    pf2 = pf                   # Penalty factors for L2 regularization
)
# Extract exposure coefficients (exclude intercept and confounders)
beta_hat <- coef(best_model) # Extract all coefficients from the fitted elastic net model
exposure_coef <- beta_hat[paste0("X", 1:p), , drop = FALSE] # Subset coefficients corresponding to exposures only

# Compute ERS in the analysis set
ERS <- as.matrix(data[analysis_ids, paste0("X", 1:p)]) %*% exposure_coef
\end{rcodebox}

In this example, the ERS is computed in the analysis set using exposure coefficients estimated from the training set. Confounders are included in model fitting but excluded from the score. Once constructed, the ERS can be used as the exposure variable in mediation analysis, for instance via the \texttt{cmest} function as the previous examples for SE-MA and PC-MA.

\subsection{Bayesian Kernel Machine Regression Causal Mediation Analysis}\label{bkmr}

Bayesian Kernel Machine Regression (BKMR) is a widely applied semiparametric model that flexibly captures the relationship between high-dimensional exposure mixtures and health outcomes, while accounting for confounding factors.~\cite{bobb2015bayesiana} BKMR is particularly suited to scenarios where the relationships between exposures, mediators, and outcomes are complex, and variable selection is essential. Additionally, BKMR offers the ability to estimate credible intervals for effect estimates, which inherently account for multiple testing, enhancing the robustness of inference.~\cite{devick2022bayesiana}

BKMR was originally developed to estimate the joint effects of exposure mixtures on the outcome. In BKMR, the outcome $Y_i$ for subject $i$ is modeled as:
\begin{equation}
    Y_i = h(\X_i) + \C_i\bbeta+\epsilon_i,\label{eq:BKMR1}
\end{equation}
where $\C_i$ represents covariates, $\epsilon_i\iid N(0,\sigma^2)$ represents independent random errors and $h$ denotes an unknown and potentially nonlinear function describing the exposure-outcome relationship.~\cite{bobb2015bayesiana,bobb2018statistical} For studies of multi-exposure mixtures, the function $h$ may include a large number of exposures of interest, and the relationship between these exposures and the health outcome can be complex, including nonlinear associations, interactions among exposures, and non-additive effects, which may lead to a high-dimensional exposure-response relationship even with a modest number of exposures. In this case, directly representing $h$ with basis functions can be challenging.~\cite{bobb2018statistical} To address this, BKMR employs a kernel machine representation, which replaces explicit basis functions with a kernel function $K(\cdot,\cdot)$ that quantifies the similarity between different exposures.~\cite{cristianini2000introduction} Liu et al. demonstrated that this representation allows the model~\eqref{eq:BKMR1} to be expressed as a mixed model:~\cite{liu2007semiparametric}
\begin{align*}
    Y_i&\sim N(h_i+\C_i\bbeta,\sigma^2), \;\text{independently, for }i=1,\ldots,n,\\
    \bm{h}&=(h(\X_1),\ldots,h(\X_n))^\top\sim N(\bm{0}, \tau\bm{K}),
\end{align*}
where $\bm{K}$ is the kernel matrix whose $(i,j)$-th entry is $K(\X_i,\X_j)$. For this paper, we focus on the Gaussian kernel, a widely used choice that leverages Euclidean distances to measure similarity between $\X_i$ and $\X_j$, where $K(\X_i, \X_j)=\exp\{-\sum_{m=1}^p(X_{im} - X_{jm})^2/\rho\}$ and $\rho$ is the tuning parameter.

Devick et al. further extended the BKMR framework to causal mediation analysis (BKMR-CMA).~\cite{devick2022bayesiana} Conceptually, BKMR-CMA aligns with the general mediation DAG in Figure~\ref{fig:mediation}. The mediator model and outcome model are written as:
\begin{gather}
    M_i = h_M(\Z_{Mi}) + \C_i\balpha_c^{\text{BKMR}} + \epsilon_{Mi},\label{eq:bkmrmed}\\ 
    Y_i = h_Y(\Z_{Yi},M_i)+\C_i\bbeta_c^{\text{BKMR}}+\epsilon_{ei},\label{eq:bkmrout}
\end{gather}
where $\epsilon_{ei}\iid N(0,\sigma_e^2)$, $\epsilon_{Mi}\iid N(0,\sigma_M^2)$, and $\epsilon_{ei}$ and $\epsilon_{Mi}$ are independent. Here, $\Z_{M}=(\X,\bm{E}_m)$ and $\Z_{Y}=(\X,\bm{E}_Y)$, where $\bm{E}_M$ and $\bm{E}_Y$ denote continuous effect modifiers in the mediator model and outcome model, respectively, if included. In addition to the mediator and outcome models, BKMR-CMA estimates the total effect of the exposure mixture on the outcome by a TE model:
\begin{equation}\label{eq:bkmrte}
    Y_i=h_{TE}(\Z_{Yi})+\C_i\bm{\gamma}_c^{\text{BKMR}}+\epsilon_{TEi},
\end{equation}
where $\epsilon_{TEi}\iid N(0,\sigma_{TE}^2)$. 

The mediator, outcome, and total effect models are then utilized to estimate the causal mediation effects via MCMC sampling. These models operate under the assumption of a mean level of confounders, with counterfactual outcomes predicted based on posterior predictive distributions. To assess the change in exposure mixture from a reference level $\X=\xxs$ to a comparative level $\X=\x$, we define $\z_M^\ast=(\xxs,\bm{e}_M)$ and $\z_Y^\ast=(\xxs,\bm{e}_Y)$. For each MCMC iteration $j=1,\ldots,J$, the process involves the following steps:
\begin{enumerate}
    \item Generate $K$ samples of the mediator $M(\z_M^\ast)$ by the fitted mediator model with all the covariates at their mean level.
    \item For each iteration, use each sample of $M(\z_M^\ast)$ to estimate the corresponding outcome of $Y(\z_Y,M(\z_M^\ast))$ and $Y(\z_Y^\ast,M(\z_M^\ast))$ from the outcome model.
    \item Calculate the $j$-th posterior sample of $Y(\z_Y,M(\z_M^\ast))$ and $Y(\z_Y^\ast,M(\z_M^\ast))$ by taking the mean of the $K$ samples of $Y(\z_Y,M(\z_M^\ast))$ and $Y(\z_Y^\ast,M(\z_M^\ast))$, respectively.
    \item Obtain the $j$-th posterior sample of $Y(\z_Y)$ and $Y(\z_Y^\ast)$ directly from the TE model with all the covariates at their mean level.
    \item Calculate the $j$-th posterior sample of TE, NDE, and NIE as
    \begin{align*}
        \widehat{TE}^{(j)} &= Y(\z_Y)^{(j)} - Y(\z_Y^\ast)^{(j)}\\
        \widehat{NDE}^{(j)} &= Y(\z_Y,M(\z_M^\ast))^{(j)} - Y(\z_Y^\ast,M(\z_M^\ast))^{(j)}\\
        \widehat{NIE}^{(j)} &= \widehat{TE}^{(j)} - \widehat{NDE}^{(j)}
    \end{align*}
\end{enumerate}

This iterative procedure provides posterior distributions for TE, NDE, and NIE, allowing for inference on the mediated effects of the exposure mixture. By leveraging the flexibility of MCMC, BKMR-CMA offers a rigorous and computationally robust framework for causal mediation analysis in the presence of complex exposure-response relationships. This framework corresponds conceptually to the general DAG for the exposure mixture introduced in Figure~\ref{fig:mediation}, where the exposure mixture $\X$ influences the mediator $M$, the outcome $Y$, and is confounded by covariates $\C$.

Since exposure mixtures often contain highly correlated variables, it is often helpful to identify exposures that contribute most strongly to the outcome while accounting for correlation structure. Bobb et al. proposed two strategies for variable selection in BKMR: component-wise variable selection and hierarchical variable selection.~\cite{bobb2015bayesiana} Component-wise variable selection uses an augmented Gaussian kernel to assign individual weights to each exposure dimension, defined as $K(\X_i,\X_j;\bm{r})=\exp\left(-\sum_{k=1}^pr_k(X_{ik}-X_{jk})^2\right)$, where $\bm{r}=(r_1,\ldots,r_p)^\top$ contains weights for each exposure. To determine these weights, a spike-and-slab prior is imposed:
\begin{equation}
    \begin{aligned}
        r_k\given \delta_k&\sim \delta_kf_1(r_k)+(1-\delta_k)P_0,\quad k=1,\ldots,p,\\
        \delta_k&\sim \text{Bernoulli}(\pi), \label{eq:BKMR2}
    \end{aligned}
\end{equation}
where $\delta_k$ indicates whether the $k$-th exposure is included in the model, $f_1$ denotes a probability density function with support on $\mathbb{R}^+$, and $P_0$ is the density with a point mass at 0. It allows the model to select exposures with significant contributions while shrinking less relevant ones toward zero. Notably, the posterior mean of $\delta_k$, known as the posterior inclusion probability (PIP), estimates the probability that the $k$-th exposure is included in the model, providing a direct measure of its importance in the mixture.

While component-wise selection helps identify influential exposures, it can struggle when exposures are highly correlated, making it difficult to determine the relative importance of individual exposures and potentially leading to suboptimal selection. An alternative approach suited for this situation is hierarchical variable selection, which incorporates the structure of the exposure mixture into the model. As mentioned in Section~\ref{mediation}, we assume exposures $X$ can be partitioned into groups $(X)_1,\ldots,(X)_g$, where within-group correlation is high while between-group correlation is low. We then assume that the indicator variables from the spike-and-slab prior in~\eqref{eq:BKMR2} are modeled with a group-level structure:
\begin{equation}
    \begin{aligned}
        \bm{\delta}_m\given\omega_m&\sim \text{Multinomial}(\omega_m,\bm{\pi}_m),\quad m=1,\ldots,g,\\
        \omega_m&\sim \text{Bernoulli}(\pi),
    \end{aligned}
\end{equation}
where $\bm{\delta}_m=\{\delta_i:X_i\in(X)_m\}$ is the vector of inclusion indicators for group $(X)_m$ and $\bm{\pi}_m$ is the corresponding vector of prior inclusion probabilities for the exposures within group $(X)_m$. The parameter $\omega_m$ determines whether the group contributes to the model, while the multinomial distribution ensures that at most one component from each group is selected at a time. This approach simplifies the model structure by focusing on representative variables from each group, effectively addressing the issue of high within-group correlation.

Hierarchical selection restricts each group to at most one selected variable, which implicitly assumes that exposures within the same group do not exert independent or interactive effects on the outcome. This assumption may be reasonable when within-group correlations are strong, as such effects are generally unidentifiable. However, in many real-world scenarios, exposures within the same group may still have distinct effects, and forcing them to share a common selection indicator could obscure important heterogeneity. As a result, researchers should be cautious when applying this method and consider whether the assumed grouping structure is appropriate for their data. Detailed priors and default settings for BKMR models, including hierarchical selection, are available in the works of Bobb et al.~\cite{bobb2015bayesiana,bobb2018statistical}.

After fitting the BKMR models, one can evaluate the importance of each exposure using PIP. The function \texttt{ExtractPIPs} from the \texttt{bkmr} package extracts these probabilities, summarizing the extent to which each exposure is selected across the MCMC iterations.~\cite{bkmrpac} Higher PIP values indicate stronger evidence that an exposure contributes to the exposure-response function $h$. In addition to evaluating variable importance using PIPs, researchers can explore the functional form of individual exposure-response relationships using the \texttt{PredictorResponseUnivar} function from the \texttt{bkmr} package. This function returns a data frame containing the exposure name, exposure values, and posterior mean and standard deviation estimates of $h$, which can be used to visualize the estimated marginal exposure-response function. These visualizations help identify potential nonlinear trends and assess how each exposure contributes to the outcome while holding other exposures constant.

With the BKMR framework in place, we can estimate causal mediation effects by fitting separate models for the mediator, outcome, and total effect. Then causal effects can be estimated using the \texttt{mediation.bkmr} function from the \texttt{causalbkmr} package.~\cite{causalbkmr} 

\subsubsection{Code for BKMR-CMA}

The following code illustrates how to implement BKMR-CMA with component-wise and hierarchical variable selection, as well as estimate total, direct, and indirect effects, using the \texttt{bkmr} package and the \texttt{causalbkmr} package in R.

First, we fit the outcome, mediator, and TE models using the \texttt{kmbayes} function from the \texttt{bkmr} package. Below is an example code snippet for fitting the outcome model with component-wise variable selection:
\begin{rcodebox}
# Fit the BKMR outcome model with component-wise variable selection
fit.y <- bkmr::kmbayes(
  y = Y,               # Outcome variable
  Z = cbind(X, M),     # Combined exposures and mediators
  X = C,               # Confounders
  iter = 10000,        # Number of MCMC iterations
  verbose = TRUE,      # Print progress updates
  varsel = TRUE        # Enable variable selection
)
\end{rcodebox}
In this code, \texttt{y} specifies the outcome variable, while \texttt{Z} refers to the exposures, mediators, and potential effect modifiers. The argument \texttt{X} includes the confounders to be adjusted for in the model. The variable selection is activated with \texttt{varsel = TRUE}, allowing the identification of significant predictors. For binary outcomes, an additional argument \texttt{family = "binomial"} can be specified to fit a logistic BKMR model. When fitting the mediator and TE models, the specifications for \texttt{y}, \texttt{Z}, and \texttt{X} should be modified accordingly.

Beyond these basic arguments, \texttt{kmbayes} offers several options for further customization. The argument \texttt{id} can be specified to include a random intercept. If new predictor values are available, \texttt{Znew} can be used to obtain posterior predictions for the function $h$. The argument \texttt{ztest} allows specifying which variables should be subject to variable selection while forcing others into the model. The \texttt{knots} argument enables the implementation of the Gaussian predictive process to approximate the kernel function in large datasets. The \texttt{rmethod} argument controls how $\bm{r}$ are sampled, with options to allow separate values for each predictor, enforce equality across predictors, or fix them to initial values. The argument \texttt{est.h} determines whether subject-specific posterior samples of $h$ should be drawn during model fitting, which increases computational time. Additionally, \texttt{starting.values} and \texttt{control.params} allow users to set initial values and tuning parameters for the MCMC algorithm.

Alternatively, to implement hierarchical variable selection, group indicators must be specified for the exposures. These can be defined based on hierarchical clustering of the exposure correlation matrix. Below is an example code snippet for fitting the outcome model:
\begin{rcodebox}
# Calculate the correlation matrix and perform hierarchical clustering
cor_mat <- cor(X, method = "pearson") # Calculate the correlation matrix for exposures
hc <- hclust(as.dist(1 - cor_mat))    # Cluster exposures based on correlation
groups <- cutree(hc, k = 3)           # Assign exposures to 3 groups based on clustering

# Fit the BKMR outcome model with hierarchical variable selection
fit.y <- bkmr::kmbayes(
  y = Y,                # Outcome variable
  Z = cbind(X, M),      # Combined exposures and mediators
  X = C,                # Confounders
  iter = 10000,         # Number of MCMC iterations
  verbose = TRUE,       # Print progress updates
  varsel = TRUE,        # Enable variable selection
  groups = c(groups, 4)) # Group assignments (mediator assigned as its own group)
\end{rcodebox}
In this code, \texttt{groups} specifies the hierarchical grouping of exposures, which is determined through hierarchical clustering. The number of clusters (\texttt{k = 3}) is chosen arbitrarily for this example, but in practice, it should be based on the clustering results and the underlying structure of the exposure mixture. The mediator $M$ is assigned as its own group to prevent its effects from being confounded with those of the exposures. As with component-wise selection, the same procedure applies to the mediator and TE models, with appropriate changes to the model inputs.

After fitting the three BKMR models, we use the \texttt{mediation.bkmr} function from the \texttt{causalbkmr} package to estimate TE, NDE, and NIE for a specified contrast in exposure levels:
\begin{rcodebox}
# Define the MCMC iterations to be used for inference
sel <- seq(5001, 10000, by = 1) # Use iterations from 5001 to 10000

# Specify exposure levels to compare (e.g., 25th vs. 75th percentile)
astar <- apply(X, 2, quantile, probs = 0.25) # Reference level
a <- apply(X, 2, quantile, probs = 0.75) # Comparative level

# Specify covariates used for predicting M and Y (e.g., mean of confounders)
X.predict <- matrix(colMeans(C), nrow = 1)
\end{rcodebox}
\begin{rcodebox}

# Estimate TE, NDE, and NIE for a change in exposures from reference to comparative level
mediation.effects <- causalbkmr::mediation.bkmr(
  a = a,                    # Comparative exposure level
  astar = astar,            # Reference exposure level
  e.y = NULL,               # Continuous effect modifier in the outcome model (set to NULL if not used)
  e.M = NULL,               # Continuous effect modifier in the mediator model (set to NULL if not used)
  fit.m = fit.m,            # Fitted BKMR mediator model
  fit.y = fit.y,            # Fitted BKMR outcome model
  fit.y.TE = fit.y.TE,      # Fitted BKMR total effect model
  X.predict.M = X.predict,  # Covariates to use for predicting the mediator
  X.predict.Y = X.predict,  # Covariates to use for predicting the outcome
  m.quant = c(0.1, 0.25, 0.5, 0.75),  # Quantiles of the mediator at which the controlled direct effect (CDE) is estimated
  alpha = 0.05,             # Significance level for confidence intervals (default 95% CI)
  sel = sel,                # Selected MCMC iterations for inference
  K = 50,                   # Number of mediator samples drawn per MCMC iteration
  seed = 123                # Random seed for reproducibility
)
\end{rcodebox}
In this function, \texttt{sel} specifies the MCMC iterations used for inference. Since the model was previously run for 10,000 iterations, the second half of the samples are selected for stability. The arguments \texttt{astar} and \texttt{a} define the reference and comparative exposure levels, respectively. If effect modifiers are included in the model, their levels in the outcome and mediator models can be specified using \texttt{e.y} and \texttt{e.M}. The inputs \texttt{fit.m}, \texttt{fit.y}, and \texttt{fit.y.TE} are the model outputs from fitting \texttt{kmbayes} to the mediator, outcome, and TE models, respectively. 

The function also requires specifying the covariates used for predicting the mediator (\texttt{X.predict.M}) and outcome (\texttt{X.predict.Y}). A common practice is to use the mean values of the baseline covariates. Additionally, \texttt{m.quant} allows users to set quantiles of the mediator for estimating the CDE. By default, \texttt{m.quant} specifies a range of quantiles, but if users want to estimate the CDE at specific mediator values, they can use the \texttt{m.value} argument to specify exact values rather than quantiles. The confidence level is controlled by \texttt{alpha}. The argument \texttt{K} determines the number of mediator samples drawn per MCMC iteration. Lastly, \texttt{seed} ensures reproducibility of results.

\section{Learning From Simulations}\label{simulation}

\subsection{Simulation Settings and Data Generation}\label{datagen}

To evaluate the performance of the outlined methods, we simulate four scenarios by crossing two levels of sample size ($n$=1,000 and $n$=2,500) with two levels of indirect effect strength, corresponding to mediator model $R^2_M$ values of $0.1$ and $0.4$. For each scenario, we generate 100 datasets. The data generation models for outcome and mediator are structured as:
\begin{align*}
    [Y_i\given M_i, \X_i, \C_i] &\sim N(M_i\beta_m+\X_i\bbeta_x+\C_i\bbeta_c, \sigma_e^2)\\
    [M_i\given \X_i, \C_i] &\sim N(\X_i\balpha_x+\C_i\balpha_c, \sigma_m^2)\\
    [\X_i\given \C_i] &\sim N(\C_i\bm{\theta}_c, \Sigma_X)\\
    \C_i &\sim N(0, \Sigma_C)
\end{align*}
where $Y_i$ is the continuous outcome for subject $i$ and $M_i$ is a single continuous mediator. The vector $\C_i$ consists of five confounders, which follow a multivariate normal distribution with a correlation structure where each confounder has a pairwise correlation of $0.2$ with the others. Additionally, an intercept term is included in $\C_i$ as a covariate in the models, though it is not explicitly shown in the equation. The exposure vector $\X_i$ consists of 30 standardized exposures, grouped into three mixture components of sizes, 5, 10, and 15, respectively. Each exposure component is generated with a predefined within-group correlation of $0.4$ (moderate), $0.8$ (strong), or $0.1$ (weak), representing different levels of dependency among exposures in the mixture. Both exposures and confounders are assumed to have unit variance, meaning that $\Sigma_X$ and $\Sigma_C$ correspond to the correlation matrices of the exposures and confounders, respectively.

Each simulated dataset includes either 1,000 or 2,500 observations, depending on the scenario. Non-zero effects are assigned selectively among the exposures. The exposure-outcome coefficients ($\bbeta_x$) are set to $0.3$ for the first of every three exposures, with the remaining exposures set to zero. The exposure-mediator effects ($\balpha_x$) are specified such that only the first three exposures of every ten exposures contribute non-zero effects, set as $0.3$, $0.6$, and $0.9$, with the remaining seven exposures having zero effects. The mediator-outcome effect is fixed at $\beta_m=0.4$. The variance parameters $\sigma_m^2$ and $\sigma_e^2$ are chosen to achieve target $R^2$ values of $R^2_Y=0.3$ for the outcome model and $R^2_M=0.1$ or $0.4$ for the mediator model, depending on the scenario. Confounders influence both the mediator and outcome, with effects set to 1 for $\balpha_c$ and $\bbeta_c$, while their effect on exposures, represented by $\theta_c$, is set to $0.1$. To simplify the model, we assume no exposure-mediator interactions.

Figure~\ref{fig:corplot} presents the correlation matrix from one simulated dataset under the scenario with $n=2,500$ and $R^2_M=0.4$. The predefined group structure among the three exposure components is clearly visible. The correlations between exposures, the mediator, and the outcome illustrate the strength of their associations. Notably, in mixture component groups with weaker internal correlations, exposures with non-zero causal effects exhibit stronger correlations with the mediator and outcome compared to those with no effects.

\subsection{Performance Metrics}

To evaluate the method performance across the four scenarios, we compute three key metrics: relative bias, true positive rate (TPR), and false positive rate (FPR). These metrics assess the accuracy of global NIE estimation and the ability to identify active exposures (those with non-zero true indirect effects) while avoiding false discoveries among null exposures (those with zero true indirect effects).

Relative bias is defined as
$$\text{Relative Bias}=\frac{\hat{\text{NIE}}-\text{NIE}_{\text{true}}}{\text{NIE}_{\text{true}}}\times100\%,$$
where $\hat{\text{NIE}}$ denotes the estimated global NIE and $\text{NIE}_{\text{true}}$ is the corresponding true value. We evaluate relative bias for three methods: SE-MA, PC-MA, and ERS-MA with the main effect only. For SE-MA, the true global NIE can be directly computed from the simulation parameters. For PC-MA and ERS-MA, which involve dimension reduction or penalized regression steps, we estimate $\text{NIE}_{\text{true}}$ empirically by fitting the respective models to a large reference dataset with 100,000 observations generated under the same data-generating mechanism. This large-sample estimate serves as a proxy for the population-level truth.

To evaluate variable selection performance, we compute the TPR and FPR as
\begin{align*}
    \text{TPR} &=\frac{\text{\# of active exposures correctly identified}}{\text{\# of active exposures}},\\
    \text{FPR} &=\frac{\text{\# of null exposures incorrectly identified as active}}{\text{\# of null exposures}}.
\end{align*}
We calculate TPR and FPR for two methods: SE-MA and BKMR-CMA. For SE-MA, we evaluate each exposure individually and declare an exposure as active if the false discovery rate (FDR) associated with its NIE estimate is less than 0.05. For BKMR-CMA, we examine the PIPs of exposures derived from the mediator model under both component-wise and hierarchical variable selection using the \texttt{ExtractPIPs} function in the \texttt{bkmr} package. In the component-wise approach, the PIP for each exposure is defined as the proportion of MCMC iterations in which that exposure is included in the regression function. In the hierarchical approach, exposures are first grouped based on their correlations, and the PIP for each exposure is computed as the product of the group-level PIP and the conditional PIP within the group. An exposure is considered selected if its PIP exceeds a specified threshold. We report results using three threshold values: 0.1, 0.3, and 0.5, which reflect varying levels of selection stringency.

Together, these metrics provide a comprehensive picture of how each method performs in estimating mediation effects and identifying key exposures under varying sample sizes and levels of mediation strength.

\subsection{Performance Comparison Across Methods}

\subsubsection{Relative Bias}

Figure~\ref{fig:sim_rbias} presents the relative bias in estimating the global NIE across all methods and simulation scenarios. Among the approaches, the unadjusted SE-MA performs the worst, yielding severely inflated estimates regardless of sample size or effect strength. When the indirect effects are weak ($R^2_M = 0.1$), the average relative bias exceeds 345\%, and increases to nearly 400\% under stronger effects ($R^2_M = 0.4$). This pattern reflects severe structural misspecification when co-exposures are omitted, particularly as the true mediated effect becomes larger. In contrast, the adjusted version of SE-MA, which includes all other exposures as covariates, substantially reduces relative bias to below 25\% across all scenarios, and below 10\% under high sample size and strong effects.

The PC-MA approach shows moderate bias, with results highly strongly influenced by the number of principal components retained. Using only the first principal component yields relatively stable biases between 103\% and 121\%, showing limited sensitivity to effect strength or sample size. This occurs because the top PC maximizes variance explained, but does not necessarily align with the mediator-related signal, leading to systematic under- or over-estimation. Including the top three PCs substantially increases bias, with relative biases ranging from 207\% to 293\%, and noticeably wider uncertainty intervals. Retaining enough PCs to explain 80\% of the variance leads to intermediate performance, with relative biases between 127\% and 164\%, better than top three PCs but still worse than using the first PC alone. These findings suggest that while adding PCs may help recover weak or diffuse signals, doing so often increases estimation noise, especially when the true signal is concentrated in lower-variance directions.

The ERS-MA approach outperforms both unadjusted SE-MA and PC-MA across all scenarios. Under strong mediation, average relative bias is as low as 18\% for $n=1{,}000$ and 11\% for $n=2{,}500$. Under weak effects, the average relative bias increases modestly to 29\% and 18\% respectively. These findings highlight the robustness of the ERS-MA approach when the estimated score captures the underlying exposure-mediator-outcome structure well. However, like other summary-based methods, its performance relies on projection and may underperform when important patterns lie outside the constructed score.

\subsubsection{TPR and FPR}

Figure~\ref{fig:sim_tpr} and Figure~\ref{fig:sim_fpr} summarize the average true positive rate (TPR) and false positive rate (FPR) across 100 simulated datasets for detecting signal exposures across the four simulation scenarios. The unadjusted SE-MA approach consistently achieves the highest TPRs, with values ranging from 0.60 under weak effects and $n=1{,}000$ to nearly 1.00 when both effect size and sample size are large. However, this sensitivity comes at the cost of inflated false discoveries: FPR ranges from 0.34 to 0.59 across scenarios, indicating frequent misclassifications of null exposures. These results reflect the lack of adjustment for co-exposures, leading to overestimation of indirect effects. In contrast, adjusted SE-MA sharply reduces FPR to below 0.02 in all settings, showing excellent specificity. However, its sensitivity suffers, especially when the sample size is small or effects are weak. Under $R^2_M = 0.1$ and $n = 1{,}000$, the average TPR is only 0.03, though it improves to 0.73 when $n = 2{,}500$ and $R^2_M = 0.4$. This is because conditioning on other correlated co-exposures leaves only the exposure's unique residual signal, which both weakens partial effects and inflates standard errors due to multicollinearity, thereby reducing power.

BKMR-CMA offers more balanced trade-offs depending on the variable selection strategy and PIP threshold. For component-wise selection, TPR generally increases with stronger indirect effects and more permissive thresholds, but does not consistently improve with larger sample size. At the 0.1 threshold, TPR under strong effects reaches 52\% for $n=1,000$ and 41\% for $n=2,500$, while FPR remains high in both cases (42\% and 36\%, respectively). Under weak effects, TPR decreases to 27\% ($n=1,000$) and 19\% ($n=2,500$), suggesting limited power in smaller effect settings despite lenient thresholds. Tightening the threshold to 0.3 or 0.5 reduces both TPR and FPR substantially, with TPRs dropping below 20\% and FPRs under 17\% in all cases.

Hierarchical selection exhibits more conservative behavior. At threshold 0.1 and strong effects, average TPRs are 24\% ($n=1,000$) and 19\% ($n=2,500$), notably lower than component-wise selection. However, FPRs are also reduced, averaging 19\% and 18\% respectively. Under weak effects, hierarchical TPRs remain below 20\% across all settings, and FPRs are controlled below 14\%. As thresholds increase to 0.3 and 0.5, TPRs decline further (e.g., 9.9\% at 0.3 and $n=2,500$), while FPRs drop below 7\%. 

These results suggest that while unadjusted SE-MA achieves the highest sensitivity, it lacks specificity. Adjusted SE-MA performs well in both dimensions when adequately powered. BKMR-CMA provides flexible trade-offs, with hierarchical selection offering better control of false positives in highly correlated mixtures, but tends to sacrifice sensitivity, especially when effects are weak.

\section{Analysis of PROTECT Dataset}\label{protect}

\begin{figure}[!htb]
    \centering
    \includegraphics[width=0.8\linewidth]{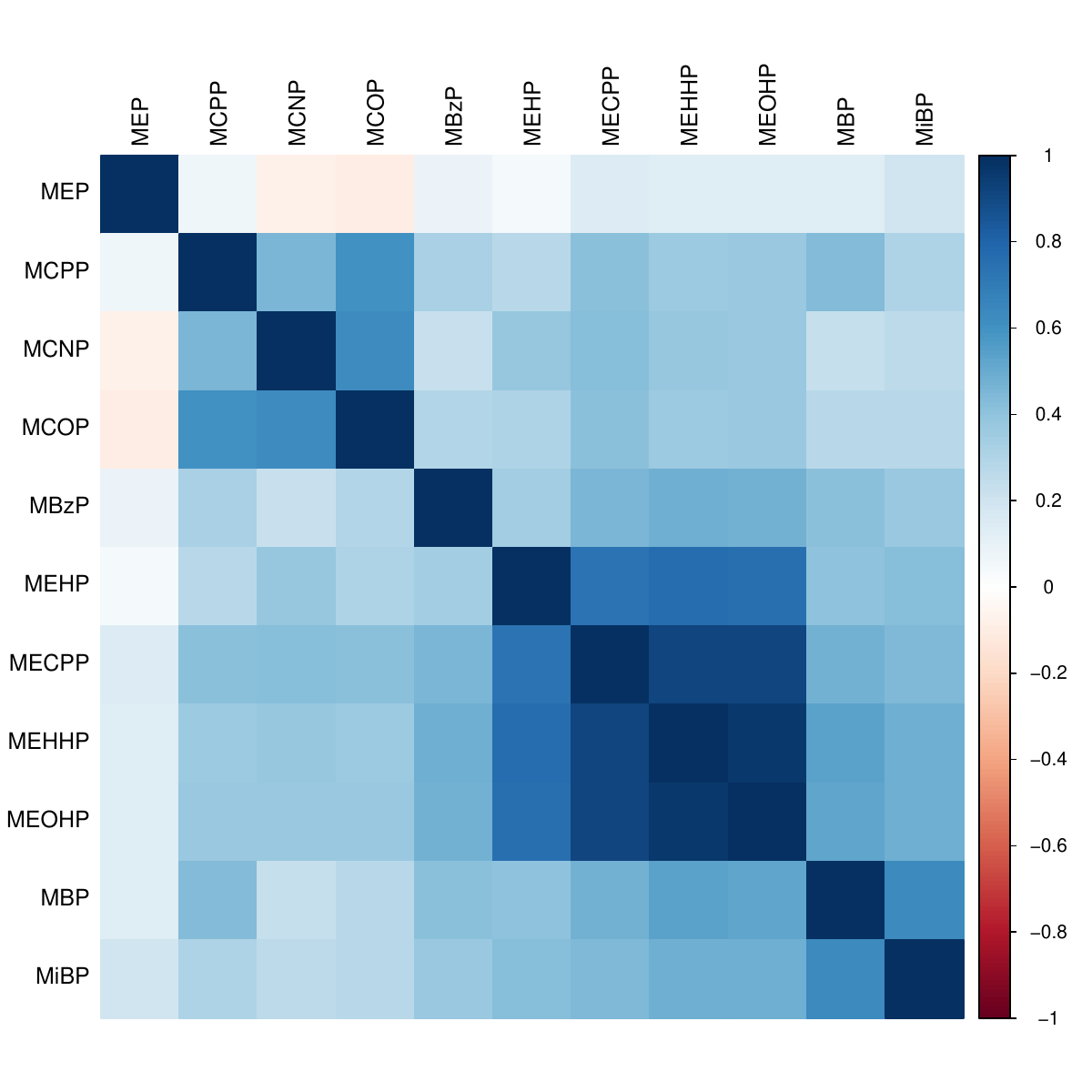}
    \caption{Spearman correlation plot of the 11 phthalate exposures in the PROTECT dataset.}
    \label{fig:data_corr}
\end{figure}

In this section, we investigate the mediation pathway from prenatal phthalate exposure to neonatal head circumference Z-score via Leukotriene E4 (LTE4), using data from the PROTECT cohort. The analysis includes phthalate metabolites measured during early pregnancy as exposures, a continuous outcome (head circumference Z-score adjusted for gestational age), and LTE4 as the mediator, a lipid mediator in the lipoxygenase pathway. Exposures are log-transformed, corrected for specific gravity, and standardized. LTE4 is selected due to its strong correlation with both phthalate exposures and head circumference in preliminary analyses. Three maternal characteristics are included as potential confounders: maternal age (4 categories), maternal education (3 categories), and pre-pregnancy body mass index (BMI) (3 categories). The final analytic dataset consists of $n=175$ observations with $p=11$ phthalate exposures, $q=1$ mediator, $s=3$ confounders (coded as seven dummy variables in the models), and a single outcome of interest. Figure~\ref{fig:data_corr} presents the Spearman correlation matrix for the 11 phthalates, revealing several distinct patterns consistent with shared parent compounds or similar exposure sources. To further characterize the correlation structure, we apply hierarchical clustering with complete linkage to define data-driven groupings of the phthalates (Figure~\ref{fig:bkmr_data_clust}). These groupings are later used to inform the hierarchical variable selection procedure in BKMR-CMA and provide additional support for subgroup structure within the mixture. Prior studies on phthalates and fetal head growth have reported mixed findings, including both positive and null associations with head circumference.~\cite{wolff2008prenatal,polanska2016effect, bloom2021association,suzuki2010prenatal} Mediation analysis in this context may help disentangle direct and indirect pathways linking phthalate exposures to fetal growth outcomes. Given the modest sample size ($n$=175), these analyses are primarily illustrative, and estimates, especially for small-to-moderate mediated effects, may be imprecise and should be interpreted with caution.

We first examine which phthalate exposures are identified as active---that is, exposures with evidence of a non-zero NIE. Under SE-MA, each exposure is evaluated separately, and exposures with NIE p-values less than or equal to 0.2 are considered potentially active. Without adjusting for co-exposures, MEP exhibits suggestive indirect effect (Table~\ref{tab:data_ind}), while co-exposure adjustment reveals two additional candidates: MCNP and MCPP (Table~\ref{tab:data_indco}). These differences highlight the importance of accounting for exposure correlation to avoid masking or inflating mediation signals. In PC-MA, mediation effects are estimated for five retained PCs explaining over 85\% of exposure variance (Figure~\ref{fig:pca_data_scree}), and only PC2 shows a significant NIE ($\text{P-value}\leq0.2$; Table~\ref{tab:pca_data}). The loading heatmap (Figure~\ref{fig:pca_data_heat}) indicates that PC2 contrasts MCNP, MCOP, and MCPP against other exposures, overlapping with the SE-MA signals for MCNP and MCPP, though interpretation is limited by the latent nature of PCs. In contrast, neither the main-effect nor the higher-order ERS-MA shows a significant NIE (Table~\ref{tab:ers_data}), though estimates are directionally consistent and reflect the overall mixture-level effect captured by the ERS score. Finally, in BKMR-CMA, component-wise variable selection identifies MCNP and MEP as having the highest PIPs in the mediator model (Figure~\ref{fig:bkmr_data_pip}), and hierarchical selection further highlights two influential groups: the MCNP-MCOP-MCPP cluster, where MCNP shows the highest conditional PIP, and the single-exposure group for MEP (Figure~\ref{fig:bkmr_data_hiermedpip}). In both selection strategies, the mediator itself also receives high inclusion probability in the outcome model (Figures~\ref{fig:bkmr_data_pip} and~\ref{fig:bkmr_data_hieroutpip}), supporting its role as a strong predictor of the outcome. Together, these results show broad agreement across methods in identifying MCNP and MEP as exposures with potential indirect effects, while also illustrating method-specific differences in how mixture structure is handled.

Next, we summarize global indirect effect estimates across all methods (Table~\ref{tab:nie_data}). Under SE-MA and PC-MA, point estimates are obtained by summing exposure- or PC-specific effect estimates, each defined under a 0-to-1 contrast (with confidence intervals approximated under an independence assumption). For ERS-MA, the global indirect effect is computed as the effect of shifting the ERS from the 25th to 75th percentile, and BKMR-CMA estimates the global indirect effect under a joint shift of all exposures from their 25th to 75th percentiles. Despite differences in modeling and contrast definitions, all estimates are close to 0, with 95\% confidence or credible intervals including 0, suggesting limited evidence of an overall mediated effect through LTE4. 

Taken together, these results demonstrate both areas of agreement and method-specific differences. MCNP and MEP are consistently highlighted across SE-MA, PC-MA, and BKMR-CMA, lending robustness to their indirect effects. At the global level, indirect effect estimates are close to 0 across all methods, with 95\% confidence or credible intervals including 0. BKMR-CMA estimates of controlled direct effects (CDEs) also remain stable across fixed levels of LTE4 (Figures~\ref{fig:bkmr_data_est} and~\ref{fig:bkmr_data_hier_est}), further supporting the limited role of mediation in this setting.

Differences across methods largely reflect how the mixture is represented and modeled, which in turn determines the estimand. SE-MA evaluates each exposure separately and does not adjust for co-exposure correlation unless explicitly modified. PC-MA reduces dimensionality by capturing major axes of variation but yields latent components that are difficult to interpret. ERS-MA compresses multiple exposures into a summary score and offers flexibility in modeling strategy, though performance depends on how the score is constructed. BKMR-CMA allows for nonlinear and interactive effects, and supports both variable- and group-level selection, albeit at higher computational cost. Overall, these findings suggest limited evidence for mediation through LTE4 and highlight the importance of applying multiple complementary methods to assess indirect effects in complex exposure settings.

\section{Discussion}\label{discussion}

\begin{figure}[!hbt]
    \centering
    \includegraphics[width=0.8\linewidth]{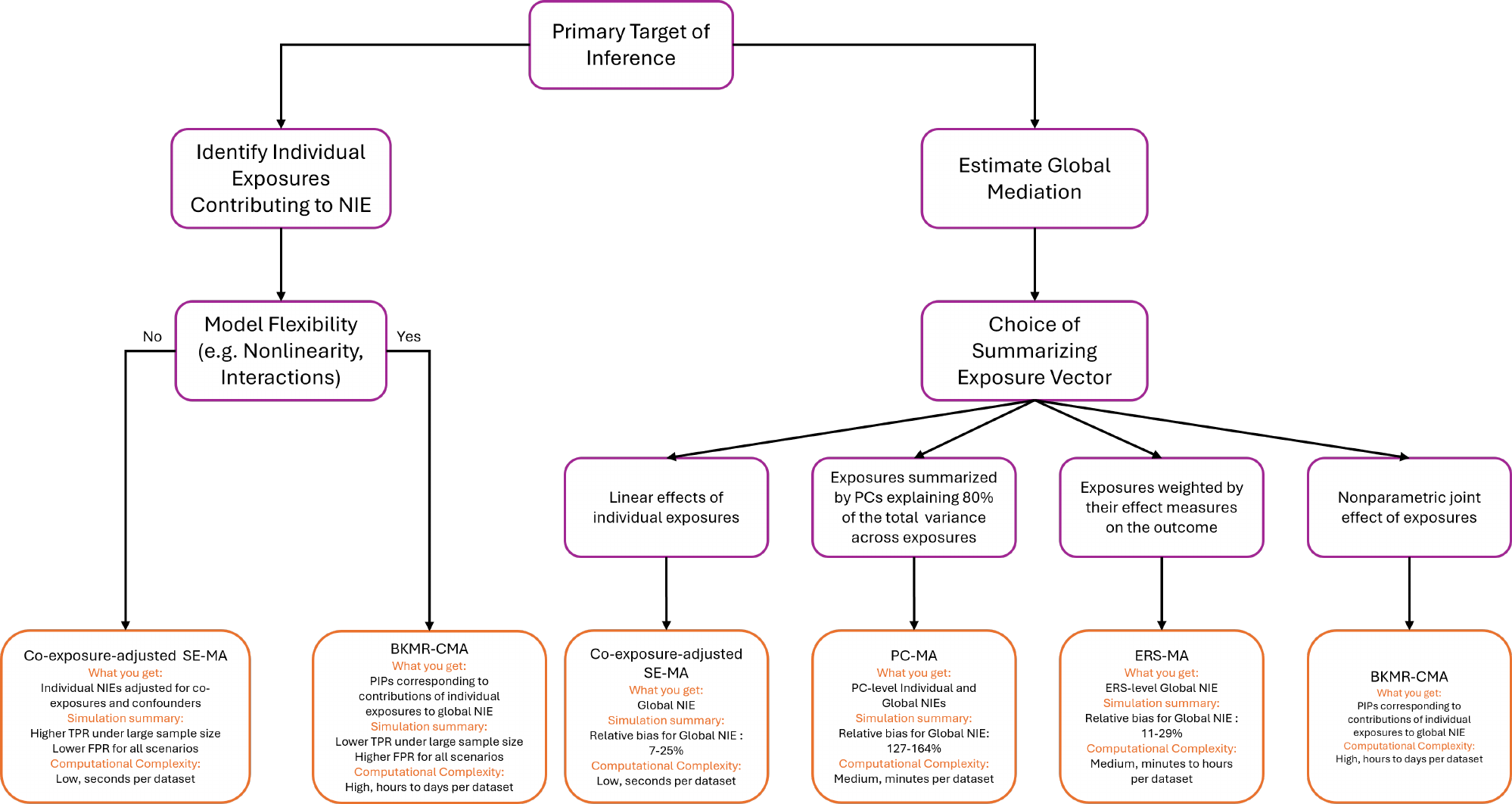}
    \caption{Practice-oriented decision framework for choosing mediation methods with correlated exposure mixtures. Estimands are defined with respect to the exposure representation adopted and should be interpreted accordingly.}
    \label{fig:decision_tree}
\end{figure}

This paper compares four strategies for conducting mediation analysis with correlated exposure mixtures: single exposure mediation analysis (SE-MA), principal component-based mediation analysis (PC-MA), environmental risk score-based mediation analysis (ERS-MA), and Bayesian kernel machine regression causal mediation analysis (BKMR-CMA). Although all four aim to quantify mediation, they reflect different causal structures and target different estimands. For SE-MA, we illustrate two possible structures depending on how co-exposures are handled. If co-exposures that affect the mediator or the outcome and are associated with the exposure of interest are omitted, they act as uncontrolled confounders of the mediator-outcome relationship, leading to biased and causally uninterpretable estimates. If they are included as covariates, SE-MA provides causally interpretable estimation of the indirect effect at the level of individual exposures, provided that the assumption of no unmeasured confounding holds for the paths of interest. Importantly, alternative causal structures are possible: if some co-exposures lie on the causal pathway, are affected by the exposure of interest, or serve as colliders, adjusting for them can introduce bias. Thus, the adjustment set must be tailored to the application and justified with a DAG and subject-matter knowledge so that the no-unmeasured-confounding assumption is plausibly satisfied.

In settings with multiple, highly correlated co-exposures, approaches that reduce dimensionality or model the mixture jointly provide practical complements to SE-MA. PC-MA replaces the exposure mixture with a lower-dimensional set of principal components, so causal interpretation applies at the component level rather than for any individual exposure. ERS-MA constructs a supervised score that summarizes the mixture into a single scalar, yielding causal effects for the score but not for the original exposures. BKMR-CMA models the joint effect of the full exposure mixture within a flexible regression framework, allowing causal interpretation at the mixture level while also enabling variable- or group-level selection to highlight influential exposures. We include assumed causal DAGs to make explicit what each approach conditions on, what is summarized, and which paths are identified.

Beyond these methodological differences, a broader point concerns the level at which mediation is assessed. Some methods yield coefficients for specific exposures, while others collapse the mixture into global summaries. In practice, these choices can lead to different scientific conclusions. In our real data analysis, SE-MA and BKMR-CMA consistently highlighted MCNP and MEP as exerting significant indirect effects via LTE4. Yet when the mixture was summarized into principal components or collapsed into an ERS, no significant indirect effects were detected. This contrast illustrates that exposures may act through distinct biological pathways that become visible only when exposures are analyzed individually, whereas treating the mixture jointly can mask those signals and yield only global mixture effects.

\begin{sidewaystable}
    \centering    
    \footnotesize
    \renewcommand{\arraystretch}{1.2}
    \caption{A summary of mediation analysis methods for exposure mixtures.}
    \setlength{\tabcolsep}{4pt}
    \begin{tabular}{|p{3.5cm}|p{1.7cm}|p{1.3cm}|p{1.6cm}|p{1.3cm}|p{1.1cm}|p{1cm}|p{3cm}|p{3.5cm}|}
        \hline
        \rowcolor{gray!30}
        \textbf{Method} & \textbf{R package} & \textbf{Handles Correlation} & \textbf{Considers Nonlinearity} & \textbf{Variable Selection} & \textbf{Individual Causal Effects} & \textbf{Global Causal Effect} & \textbf{Causal Interpretability} & \textbf{Computational Complexity (Typical Runtime)}\\
        \hline
        SE-MA without co-exposure & CMAverse~\cite{shi2021cmaverse} & No & No & No & No & No & Not causal because of omitted confounding & Low  (seconds to minutes per dataset)\\
        \rowcolor{gray!10}
        SE-MA with co-exposure & CMAverse~\cite{shi2021cmaverse} & No & No & No & Yes & No & Exposure-level, if assumptions hold & Low  (seconds to minutes per dataset)\\
        PC-MA & stats (prcomp)~\cite{R}, CMAverse~\cite{shi2021cmaverse} & Yes & No & No & No & Yes & PC level; not exposure level & Medium (minutes per dataset, depends on exposure count)\\
        \rowcolor{gray!10}
        ERS-MA (elastic net, main effects only)~\cite{wang2018associations,park2014environmentala,park2017constructiona} & gcdnet~\cite{gcdnet}, glmnet~\cite{glmnet}, CMAverse~\cite{shi2021cmaverse} & Yes & No & Yes & No & Yes & ERS level; not exposure level & Medium (minutes to hours, depends on regularization tuning)\\
        ERS-MA (elastic net, main effects, squared terms and interactions)~\cite{wang2018associations,park2014environmentala,park2017constructiona} & gcdnet~\cite{gcdnet}, glmnet~\cite{glmnet}, CMAverse~\cite{shi2021cmaverse} & Yes & Yes & Yes & No & Yes & ERS level; not exposure level & Medium (minutes to hours, depends on regularization tuning) \\
        \rowcolor{gray!10}
        BKMR-CMA~\cite{bobb2015bayesiana,devick2022bayesiana} & bkmr~\cite{bkmr}, causalbkmr~\cite{causalbkmr} & Yes & Yes & Yes & No & Yes & Mixture level; not individual exposure level & High (hours to days per dataset, due to MCMC sampling)\\
        \hline 
    \end{tabular}
    \label{tab:summary}
\end{sidewaystable}

The overarching message is therefore not that one method is universally superior, but that each makes explicit trade-offs between interpretability, flexibility, and the target and task of inference. Researchers should align method choice with their primary aim: to identify particular exposures acting through biological pathways, or to evaluate overall mediation at the mixture level. To support this process, we provide a decision tree (Figure~\ref{fig:decision_tree}) that maps common objectives to appropriate methods, and a summary table (Table~\ref{tab:summary}) outlining key methodological features, including their treatment of correlation, nonlinearity, and variable selection. Together, these tools are intended to clarify assumptions and estimands, and to help applied researchers select the method most consistent with their objectives and data structure.

Finally, we note two additional practical considerations for applied researchers. First, although our illustrations focused on continuous outcomes, most of the methods discussed are not restricted to that setting. For example, packages \texttt{CMAverse}, \texttt{gcdnet}, and \texttt{BKMR-CMA} support non-continuous outcomes such as binary data. Thus, if the outcome of interest is non-continuous, the same workflow can generally be applied with only minor modifications to the outcome model specification. Second, all causal mediation analyses rely on identification assumptions that require careful consideration. Two central assumptions are consistency (which may be threatened by measurement error in environmental exposures) and the absence of unmeasured confounding along the exposure–mediator–outcome paths. We recommend that researchers explicitly justify their adjustment set using a DAG and subject-matter knowledge, and conduct sensitivity analyses when possible. For example, E-values can be used to assess the robustness of results to potential unmeasured confounding, while other sensitivity analysis approaches may be needed to evaluate the impact of measurement error.~\cite{vanderweele2017sensitivity}

\FloatBarrier

\section*{Acknowledgments}

The authors gratefully acknowledge the support of the Yale School of Public Health in the development of this work.

\section*{Conflict of interest}

The authors declare no potential conflict of interests.

\section*{Data Availability Statement}

The data that support the findings of this study are openly available in GitHub at \url{https://github.com/ysph-dsde/Mixture-Mediation-Tutorial}.

\newpage

\bibliographystyle{unsrtnat}
\bibliography{ref}

\begin{thebibliography}{61}
\providecommand{\natexlab}[1]{#1}
\providecommand{\url}[1]{\texttt{#1}}
\expandafter\ifx\csname urlstyle\endcsname\relax
  \providecommand{\doi}[1]{doi: #1}\else
  \providecommand{\doi}{doi: \begingroup \urlstyle{rm}\Url}\fi

\bibitem[VanderWeele(2016)]{vanderweele2016mediation}
Tyler~J. VanderWeele.
\newblock Mediation {{Analysis}}: {{A Practitioner}}'s {{Guide}}.
\newblock \emph{Annual Review of Public Health}, 37\penalty0 (1):\penalty0 17--32, 2016.
\newblock \doi{10.1146/annurev-publhealth-032315-021402}.

\bibitem[Dominici et~al.(2010)Dominici, Peng, Barr, and Bell]{dominici2010protecting}
Francesca Dominici, Roger~D. Peng, Christopher~D. Barr, and Michelle~L. Bell.
\newblock Protecting {{Human Health From Air Pollution}}: {{Shifting From}} a {{Single-pollutant}} to a {{Multipollutant Approach}}.
\newblock \emph{Epidemiology}, 21\penalty0 (2):\penalty0 187--194, 2010.
\newblock \doi{10.1097/EDE.0b013e3181cc86e8}.

\bibitem[Toccalino et~al.(2012)Toccalino, Norman, and Scott]{toccalino2012chemical}
Patricia~L. Toccalino, Julia~E. Norman, and Jonathon~C. Scott.
\newblock Chemical mixtures in untreated water from public-supply wells in the {{U}}.{{S}}. --- {{Occurrence}}, composition, and potential toxicity.
\newblock \emph{Science of The Total Environment}, 431:\penalty0 262--270, 2012.
\newblock \doi{10.1016/j.scitotenv.2012.05.044}.

\bibitem[Wang et~al.(2018)Wang, Mukherjee, and Park]{wang2018associations}
Xin Wang, Bhramar Mukherjee, and Sung~Kyun Park.
\newblock Associations of cumulative exposure to heavy metal mixtures with obesity and its comorbidities among {{U}}.{{S}}. adults in {{NHANES}} 2003--2014.
\newblock \emph{Environment International}, 121:\penalty0 683--694, 2018.
\newblock \doi{10.1016/j.envint.2018.09.035}.

\bibitem[Henn et~al.(2012)Henn, Schnaas, Ettinger, Schwartz, {Lamadrid-Figueroa}, {Hern{\'a}ndez-Avila}, Amarasiriwardena, Hu, Bellinger, Wright, and {T{\'e}llez-Rojo}]{henn2012associations}
Birgit~Claus Henn, Lourdes Schnaas, Adrienne~S. Ettinger, Joel Schwartz, H{\'e}ctor {Lamadrid-Figueroa}, Mauricio {Hern{\'a}ndez-Avila}, Chitra Amarasiriwardena, Howard Hu, David~C. Bellinger, Robert~O. Wright, and Martha~Mar{\'i}a {T{\'e}llez-Rojo}.
\newblock Associations of {{Early Childhood Manganese}} and {{Lead Coexposure}} with {{Neurodevelopment}}.
\newblock \emph{Environmental Health Perspectives}, 120\penalty0 (1):\penalty0 126--131, 2012.
\newblock \doi{10.1289/ehp.1003300}.

\bibitem[Claus~Henn et~al.(2014)Claus~Henn, Coull, and Wright]{claushenn2014chemical}
Birgit Claus~Henn, Brent~A. Coull, and Robert~O. Wright.
\newblock Chemical mixtures and children's health.
\newblock \emph{Current Opinion in Pediatrics}, 26\penalty0 (2):\penalty0 223--229, 2014.
\newblock \doi{10.1097/MOP.0000000000000067}.

\bibitem[Hamra and Buckley(2018)]{hamra2018environmental}
Ghassan~B Hamra and Jessie~P Buckley.
\newblock Environmental exposure mixtures: questions and methods to address them.
\newblock \emph{Current epidemiology reports}, 5\penalty0 (2):\penalty0 160--165, 2018.
\newblock \doi{10.1007/s40471-018-0145-0}.

\bibitem[Aung et~al.(2020)Aung, Song, Ferguson, Cantonwine, Zeng, McElrath, Pennathur, Meeker, and Mukherjee]{aung2020application}
Max~T Aung, Yanyi Song, Kelly~K Ferguson, David~E Cantonwine, Lixia Zeng, Thomas~F McElrath, Subramaniam Pennathur, John~D Meeker, and Bhramar Mukherjee.
\newblock Application of an analytical framework for multivariate mediation analysis of environmental data.
\newblock \emph{Nature communications}, 11\penalty0 (1):\penalty0 5624, 2020.
\newblock \doi{10.1038/s41467-020-19335-2}.

\bibitem[Zitko(1994)]{zitko1994principal}
V.~Zitko.
\newblock Principal component analysis in the evaluation of environmental data.
\newblock \emph{Marine Pollution Bulletin}, 28\penalty0 (12):\penalty0 718--722, 1994.
\newblock \doi{10.1016/0025-326X(94)90329-8}.

\bibitem[Reid and Spencer(2009)]{reid2009use}
M.K. Reid and K.L. Spencer.
\newblock Use of principal components analysis ({{PCA}}) on estuarine sediment datasets: {{The}} effect of data pre-treatment.
\newblock \emph{Environmental Pollution}, 157\penalty0 (8-9):\penalty0 2275--2281, 2009.
\newblock \doi{10.1016/j.envpol.2009.03.033}.

\bibitem[Ma and Amos(2012)]{ma2012principal}
Jianzhong Ma and Christopher~I. Amos.
\newblock Principal {{Components Analysis}} of {{Population Admixture}}.
\newblock \emph{PLoS ONE}, 7\penalty0 (7):\penalty0 e40115, 2012.
\newblock \doi{10.1371/journal.pone.0040115}.

\bibitem[Abdi and Williams(2010)]{abdi2010principal}
Herv{\'e} Abdi and Lynne~J. Williams.
\newblock Principal component analysis.
\newblock \emph{WIREs Computational Statistics}, 2\penalty0 (4):\penalty0 433--459, 2010.
\newblock \doi{10.1002/wics.101}.

\bibitem[Park et~al.(2014)Park, Tao, Meeker, Harlow, and Mukherjee]{park2014environmentala}
Sung~Kyun Park, Yebin Tao, John~D. Meeker, Siob{\'a}n~D. Harlow, and Bhramar Mukherjee.
\newblock Environmental {{Risk Score}} as a {{New Tool}} to {{Examine Multi-Pollutants}} in {{Epidemiologic Research}}: {{An Example}} from the {{NHANES Study Using Serum Lipid Levels}}.
\newblock \emph{PLoS ONE}, 9\penalty0 (6):\penalty0 e98632, 2014.
\newblock \doi{10.1371/journal.pone.0098632}.

\bibitem[Bobb et~al.(2015)Bobb, Valeri, Claus~Henn, Christiani, Wright, Mazumdar, Godleski, and Coull]{bobb2015bayesiana}
Jennifer~F. Bobb, Linda Valeri, Birgit Claus~Henn, David~C. Christiani, Robert~O. Wright, Maitreyi Mazumdar, John~J. Godleski, and Brent~A. Coull.
\newblock Bayesian kernel machine regression for estimating the health effects of multi-pollutant mixtures.
\newblock \emph{Biostatistics}, 16\penalty0 (3):\penalty0 493--508, 2015.
\newblock \doi{10.1093/biostatistics/kxu058}.

\bibitem[Devick et~al.(2022)Devick, Bobb, Mazumdar, Claus~Henn, Bellinger, Christiani, Wright, Williams, Coull, and Valeri]{devick2022bayesiana}
Katrina~L. Devick, Jennifer~F. Bobb, Maitreyi Mazumdar, Birgit Claus~Henn, David~C. Bellinger, David~C. Christiani, Robert~O. Wright, Paige~L. Williams, Brent~A. Coull, and Linda Valeri.
\newblock Bayesian kernel machine regression-causal mediation analysis.
\newblock \emph{Statistics in Medicine}, 41\penalty0 (5):\penalty0 860--876, 2022.
\newblock \doi{10.1002/sim.9255}.

\bibitem[Aung et~al.(2021)Aung, Yu, Ferguson, Cantonwine, Zeng, McElrath, Pennathur, Mukherjee, and Meeker]{aung2021cross}
Max~T Aung, Youfei Yu, Kelly~K Ferguson, David~E Cantonwine, Lixia Zeng, Thomas~F McElrath, Subramaniam Pennathur, Bhramar Mukherjee, and John~D Meeker.
\newblock Cross-sectional estimation of endogenous biomarker associations with prenatal phenols, phthalates, metals, and polycyclic aromatic hydrocarbons in single-pollutant and mixtures analysis approaches.
\newblock \emph{Environmental health perspectives}, 129\penalty0 (3):\penalty0 037007, 2021.
\newblock \doi{10.1289/EHP7396}.

\bibitem[Robins and Greenland(1992)]{robins1992identifiability}
James~M. Robins and Sander Greenland.
\newblock Identifiability and {{Exchangeability}} for {{Direct}} and {{Indirect Effects}}:.
\newblock \emph{Epidemiology}, 3\penalty0 (2):\penalty0 143--155, 1992.
\newblock \doi{10.1097/00001648-199203000-00013}.

\bibitem[Pearl(2001)]{pearl2001direct}
Judea Pearl.
\newblock Direct and indirect effects.
\newblock In \emph{Proceedings of the Seventeenth Conference on Uncertainty in Artificial Intelligence}, UAI'01, pages 411--420, San Francisco, CA, USA, 2001. The Association for Uncertainty in Artificial Intelligence, Morgan Kaufmann Publishers Inc.
\newblock ISBN 1558608001.
\newblock \doi{10.5555/2074022.2074073}.
\newblock URL \url{https://dl.acm.org/doi/10.5555/2074022.2074073}.

\bibitem[Vanderweele and Vansteelandt(2009)]{vanderweele2009conceptual}
Tyler~J. Vanderweele and Stijn Vansteelandt.
\newblock Conceptual issues concerning mediation, interventions and composition.
\newblock \emph{Statistics and Its Interface}, 2\penalty0 (4):\penalty0 457--468, 2009.
\newblock \doi{10.4310/SII.2009.v2.n4.a7}.

\bibitem[VanderWeele and Vansteelandt(2010)]{vanderweele2010odds}
T.~J. VanderWeele and S.~Vansteelandt.
\newblock Odds {{Ratios}} for {{Mediation Analysis}} for a {{Dichotomous Outcome}}.
\newblock \emph{American Journal of Epidemiology}, 172\penalty0 (12):\penalty0 1339--1348, 2010.
\newblock \doi{10.1093/aje/kwq332}.

\bibitem[Imai et~al.(2010{\natexlab{a}})Imai, Keele, and Tingley]{imai2010generala}
Kosuke Imai, Luke Keele, and Dustin Tingley.
\newblock A general approach to causal mediation analysis.
\newblock \emph{Psychological Methods}, 15\penalty0 (4):\penalty0 309--334, 2010{\natexlab{a}}.
\newblock \doi{10.1037/a0020761}.

\bibitem[Valeri and VanderWeele(2013)]{valeri2013mediation}
Linda Valeri and Tyler~J. VanderWeele.
\newblock Mediation analysis allowing for exposure--mediator interactions and causal interpretation: {{Theoretical}} assumptions and implementation with {{SAS}} and {{SPSS}} macros.
\newblock \emph{Psychological Methods}, 18\penalty0 (2):\penalty0 137--150, 2013.
\newblock \doi{10.1037/a0031034}.

\bibitem[Ding(2024)]{ding2024first}
Peng Ding.
\newblock \emph{A first course in causal inference}.
\newblock CRC Press, 1st edition, 2024.
\newblock \doi{10.1201/9781003484080}.

\bibitem[Lange et~al.(2017)Lange, Hansen, S{\o}rensen, and Galatius]{lange2017applied}
Theis Lange, Kim~Wadt Hansen, Rikke S{\o}rensen, and S{\o}ren Galatius.
\newblock Applied mediation analyses: a review and tutorial.
\newblock \emph{Epidemiology and health}, 39:\penalty0 e2017035, 2017.
\newblock \doi{10.4178/epih.e2017035}.

\bibitem[Rubin(1990)]{rubin1990formal}
Donald~B Rubin.
\newblock Formal mode of statistical inference for causal effects.
\newblock \emph{Journal of statistical planning and inference}, 25\penalty0 (3):\penalty0 279--292, 1990.
\newblock ISSN 0378-3758.
\newblock \doi{10.1016/0378-3758(90)90077-8}.

\bibitem[VanderWeele(2015)]{vanderweele2015explanation}
Tyler~J. VanderWeele.
\newblock \emph{Explanation in Causal Inference: Methods for Mediation and Interaction}.
\newblock Oxford University Press, New York, NY, 2015.
\newblock ISBN 978-0-19-932587-0.

\bibitem[Imai et~al.(2010{\natexlab{b}})Imai, Keele, and Yamamoto]{imai2010identification}
Kosuke Imai, Luke Keele, and Teppei Yamamoto.
\newblock Identification, {{Inference}} and {{Sensitivity Analysis}} for {{Causal Mediation Effects}}.
\newblock \emph{Statistical Science}, 25\penalty0 (1):\penalty0 57--71, 2010{\natexlab{b}}.
\newblock \doi{10.1214/10-STS321}.

\bibitem[Baron and Kenny(1986)]{baron1986moderator}
Reuben~M. Baron and David~A. Kenny.
\newblock The moderator--mediator variable distinction in social psychological research: {{Conceptual}}, strategic, and statistical considerations.
\newblock \emph{Journal of Personality and Social Psychology}, 51\penalty0 (6):\penalty0 1173--1182, 1986.
\newblock \doi{10.1037/0022-3514.51.6.1173}.

\bibitem[Dunn(1961)]{dunn1961multiple}
Olive~Jean Dunn.
\newblock Multiple comparisons among means.
\newblock \emph{Journal of the American statistical association}, 56\penalty0 (293):\penalty0 52--64, 1961.
\newblock \doi{10.1080/01621459.1961.10482090}.

\bibitem[Holm(1979)]{holm1979simple}
Sture Holm.
\newblock A simple sequentially rejective multiple test procedure.
\newblock \emph{Scandinavian journal of statistics}, pages 65--70, 1979.

\bibitem[Benjamini and Hochberg(1995)]{benjamini1995controlling}
Yoav Benjamini and Yosef Hochberg.
\newblock Controlling the {{False Discovery Rate}}: {{A Practical}} and {{Powerful Approach}} to {{Multiple Testing}}.
\newblock \emph{Journal of the Royal Statistical Society Series B: Statistical Methodology}, 57\penalty0 (1):\penalty0 289--300, 1995.
\newblock \doi{10.1111/j.2517-6161.1995.tb02031.x}.

\bibitem[Shi et~al.(2021)Shi, Choirat, Coull, VanderWeele, and Valeri]{shi2021cmaverse}
Baoyi Shi, Christine Choirat, Brent~A. Coull, Tyler~J. VanderWeele, and Linda Valeri.
\newblock {{CMAverse}}: {{A Suite}} of {{Functions}} for {{Reproducible Causal Mediation Analyses}}.
\newblock \emph{Epidemiology}, 32\penalty0 (5):\penalty0 e20--e22, 2021.
\newblock \doi{10.1097/EDE.0000000000001378}.

\bibitem[VanderWeele and Vansteelandt(2014)]{vanderweele2014mediation}
Tyler VanderWeele and Stijn Vansteelandt.
\newblock Mediation {{Analysis}} with {{Multiple Mediators}}.
\newblock \emph{Epidemiologic Methods}, 2\penalty0 (1), 2014.
\newblock \doi{10.1515/em-2012-0010}.

\bibitem[Tchetgen~Tchetgen(2013)]{tchetgen2013inverse}
Eric~J Tchetgen~Tchetgen.
\newblock Inverse odds ratio-weighted estimation for causal mediation analysis.
\newblock \emph{Statistics in medicine}, 32\penalty0 (26):\penalty0 4567--4580, 2013.
\newblock \doi{10.1002/sim.5864}.

\bibitem[Vansteelandt et~al.(2012)Vansteelandt, Bekaert, and Lange]{vansteelandt2012imputation}
Stijn Vansteelandt, Maarten Bekaert, and Theis Lange.
\newblock Imputation strategies for the estimation of natural direct and indirect effects.
\newblock \emph{Epidemiologic Methods}, 1\penalty0 (1):\penalty0 131--158, 2012.
\newblock \doi{10.1515/2161-962X.1014}.

\bibitem[VanderWeele and Tchetgen~Tchetgen(2017)]{vanderweele2017mediation}
Tyler~J VanderWeele and Eric~J Tchetgen~Tchetgen.
\newblock Mediation analysis with time varying exposures and mediators.
\newblock \emph{Journal of the Royal Statistical Society Series B: Statistical Methodology}, 79\penalty0 (3):\penalty0 917--938, 2017.
\newblock \doi{10.1111/rssb.12194}.

\bibitem[Robins(1986)]{robins1986new}
James Robins.
\newblock A new approach to causal inference in mortality studies with a sustained exposure period—application to control of the healthy worker survivor effect.
\newblock \emph{Mathematical modelling}, 7\penalty0 (9-12):\penalty0 1393--1512, 1986.
\newblock \doi{10.1016/0270-0255(86)90088-6}.

\bibitem[Gibson et~al.(2019)Gibson, Nunez, Abuawad, Zota, Renzetti, Devick, Gennings, Goldsmith, Coull, and Kioumourtzoglou]{gibson2019overview}
Elizabeth~A. Gibson, Yanelli Nunez, Ahlam Abuawad, Ami~R. Zota, Stefano Renzetti, Katrina~L. Devick, Chris Gennings, Jeff Goldsmith, Brent~A. Coull, and Marianthi-Anna Kioumourtzoglou.
\newblock An overview of methods to address distinct research questions on environmental mixtures: An application to persistent organic pollutants and leukocyte telomere length.
\newblock \emph{Environmental Health}, 18\penalty0 (1):\penalty0 76, 2019.
\newblock \doi{10.1186/s12940-019-0515-1}.

\bibitem[Jolliffe and Cadima(2016)]{jolliffe2016principal}
Ian~T. Jolliffe and Jorge Cadima.
\newblock Principal component analysis: A review and recent developments.
\newblock \emph{Philosophical Transactions of the Royal Society A: Mathematical, Physical and Engineering Sciences}, 374\penalty0 (2065):\penalty0 20150202, 2016.
\newblock \doi{10.1098/rsta.2015.0202}.

\bibitem[{R Core Team}(2025)]{R}
{R Core Team}.
\newblock \emph{R: A Language and Environment for Statistical Computing}.
\newblock R Foundation for Statistical Computing, Vienna, Austria, 2025.
\newblock URL \url{https://www.R-project.org/}.

\bibitem[Boss(2023)]{boss2023shrinkage}
Jonathan Boss.
\newblock \emph{Shrinkage Methods for High-Dimensional Regression and Mediation Models}.
\newblock PhD thesis, University of Michigan, Ann Arbor, MI, 2023.

\bibitem[Kassambara and Mundt(2020)]{factoextra}
Alboukadel Kassambara and Fabian Mundt.
\newblock factoextra: Extract and visualize the results of multivariate data analyses.
\newblock \url{https://CRAN.R-project.org/package=factoextra}, 2020.
\newblock R package version 1.0.7.

\bibitem[Kaiser(1960)]{kaiser1960application}
Henry~F Kaiser.
\newblock The application of electronic computers to factor analysis.
\newblock \emph{Educational and psychological measurement}, 20\penalty0 (1):\penalty0 141--151, 1960.
\newblock \doi{10.1177/001316446002000116}.

\bibitem[Park et~al.(2017)Park, Zhao, and Mukherjee]{park2017constructiona}
Sung~Kyun Park, Zhangchen Zhao, and Bhramar Mukherjee.
\newblock Construction of environmental risk score beyond standard linear models using machine learning methods: Application to metal mixtures, oxidative stress and cardiovascular disease in {{NHANES}}.
\newblock \emph{Environmental Health}, 16\penalty0 (1):\penalty0 102, 2017.
\newblock \doi{10.1186/s12940-017-0310-9}.

\bibitem[Chipman et~al.(2010)Chipman, George, and McCulloch]{chipman2010bart}
Hugh~A. Chipman, Edward~I. George, and Robert~E. McCulloch.
\newblock {BART: Bayesian additive regression trees}.
\newblock \emph{The Annals of Applied Statistics}, 4\penalty0 (1):\penalty0 266--298, 2010.
\newblock \doi{10.1214/09-AOAS285}.
\newblock URL \url{https://doi.org/10.1214/09-AOAS285}.

\bibitem[Van~der Laan et~al.(2007)Van~der Laan, Polley, and Hubbard]{van2007super}
Mark~J Van~der Laan, Eric~C Polley, and Alan~E Hubbard.
\newblock Super learner.
\newblock \emph{Statistical applications in genetics and molecular biology}, 6\penalty0 (1), 2007.
\newblock \doi{10.2202/1544-6115.1309}.

\bibitem[Zou and Hastie(2005)]{zou2005regularization}
Hui Zou and Trevor Hastie.
\newblock Regularization and {{Variable Selection Via}} the {{Elastic Net}}.
\newblock \emph{Journal of the Royal Statistical Society Series B: Statistical Methodology}, 67\penalty0 (2):\penalty0 301--320, 2005.
\newblock \doi{10.1111/j.1467-9868.2005.00503.x}.

\bibitem[Yang et~al.(2022)Yang, Gu, and Zou]{gcdnet}
Yi~Yang, Yuwen Gu, and Hui Zou.
\newblock gcdnet: The (adaptive) lasso and elastic net penalized least squares, logistic regression, hybrid huberized support vector machines, squared hinge loss support vector machines and expectile regression using a fast generalized coordinate descent algorithm.
\newblock \url{https://CRAN.R-project.org/package=gcdnet}, 2022.
\newblock R package version 1.0.6.

\bibitem[Bobb et~al.(2018)Bobb, Claus~Henn, Valeri, and Coull]{bobb2018statistical}
Jennifer~F. Bobb, Birgit Claus~Henn, Linda Valeri, and Brent~A. Coull.
\newblock Statistical software for analyzing the health effects of multiple concurrent exposures via {{Bayesian}} kernel machine regression.
\newblock \emph{Environmental Health}, 17\penalty0 (1):\penalty0 67, 2018.
\newblock \doi{10.1186/s12940-018-0413-y}.

\bibitem[Cristianini and Shawe-Taylor(2000)]{cristianini2000introduction}
Nello Cristianini and John Shawe-Taylor.
\newblock \emph{An introduction to support vector machines and other kernel-based learning methods}.
\newblock Cambridge university press, 2000.
\newblock ISBN 9780511801389.
\newblock \doi{10.1017/CBO9780511801389}.

\bibitem[Liu et~al.(2007)Liu, Lin, and Ghosh]{liu2007semiparametric}
Dawei Liu, Xihong Lin, and Debashis Ghosh.
\newblock Semiparametric regression of multidimensional genetic pathway data: least-squares kernel machines and linear mixed models.
\newblock \emph{Biometrics}, 63\penalty0 (4):\penalty0 1079--1088, 2007.
\newblock \doi{10.1111/j.1541-0420.2007.00799.x}.

\bibitem[Bobb(2022{\natexlab{a}})]{bkmrpac}
Jennifer~F. Bobb.
\newblock bkmr: Bayesian kernel machine regression.
\newblock \url{https://CRAN.R-project.org/package=bkmr}, 2022{\natexlab{a}}.
\newblock R package version 0.2.2.

\bibitem[Chai et~al.(2025)Chai, Devick, and Valeri]{causalbkmr}
Zilan Chai, Katrina Devick, and Linda Valeri.
\newblock causalbkmr: Causal inference with bayesian kernel machine regression.
\newblock \url{https://github.com/zc2326/causalbkmr}, 2025.
\newblock R package version 0.1.0, commit 4a33c0bbe76c7b5c55604064413b9be3e379e901.

\bibitem[Wolff et~al.(2008)Wolff, Engel, Berkowitz, Ye, Silva, Zhu, Wetmur, and Calafat]{wolff2008prenatal}
Mary~S. Wolff, Stephanie~M. Engel, Gertrud~S. Berkowitz, Xiaoyun Ye, Manori~J. Silva, Chenbo Zhu, James Wetmur, and Antonia~M. Calafat.
\newblock Prenatal {{Phenol}} and {{Phthalate Exposures}} and {{Birth Outcomes}}.
\newblock \emph{Environmental Health Perspectives}, 116\penalty0 (8):\penalty0 1092--1097, 2008.
\newblock \doi{10.1289/ehp.11007}.

\bibitem[Pola{\'n}ska et~al.(2016)Pola{\'n}ska, Ligocka, Sobala, and Hanke]{polanska2016effect}
Kinga Pola{\'n}ska, Danuta Ligocka, Wojciech Sobala, and Wojciech Hanke.
\newblock Effect of environmental phthalate exposure on pregnancy duration and birth outcomes.
\newblock \emph{International Journal of Occupational Medicine and Environmental Health}, 29\penalty0 (4):\penalty0 683--697, 2016.
\newblock \doi{10.13075/ijomeh.1896.00691}.

\bibitem[Bloom et~al.(2021)Bloom, Valachovic, Begum, Kucklick, Brock, Wenzel, Wineland, Cruze, Unal, and Newman]{bloom2021association}
Michael~S. Bloom, Edward~L. Valachovic, Thoin~F. Begum, John~R. Kucklick, John~W. Brock, Abby~G. Wenzel, Rebecca~J. Wineland, Lori Cruze, Elizabeth~R. Unal, and Roger~B. Newman.
\newblock Association between gestational phthalate exposure and newborn head circumference; impacts by race and sex.
\newblock \emph{Environmental Research}, 195:\penalty0 110763, 2021.
\newblock \doi{10.1016/j.envres.2021.110763}.

\bibitem[Suzuki et~al.(2010)Suzuki, Niwa, Yoshinaga, Mizumoto, Serizawa, and Shiraishi]{suzuki2010prenatal}
Yayoi Suzuki, Mayu Niwa, Jun Yoshinaga, Yoshifumi Mizumoto, Shigeko Serizawa, and Hiroaki Shiraishi.
\newblock Prenatal exposure to phthalate esters and {{PAHs}} and birth outcomes.
\newblock \emph{Environment International}, 36\penalty0 (7):\penalty0 699--704, 2010.
\newblock \doi{10.1016/j.envint.2010.05.003}.

\bibitem[Friedman et~al.(2010)Friedman, Hastie, and Tibshirani]{glmnet}
Jerome Friedman, Trevor Hastie, and Robert Tibshirani.
\newblock Regularization paths for generalized linear models via coordinate descent.
\newblock \emph{Journal of Statistical Software}, 33\penalty0 (1):\penalty0 1--22, 2010.
\newblock \doi{10.18637/jss.v033.i01}.

\bibitem[Bobb(2022{\natexlab{b}})]{bkmr}
Jennifer~F. Bobb.
\newblock bkmr: Bayesian kernel machine regression.
\newblock \url{https://CRAN.R-project.org/package=bkmr}, 2022{\natexlab{b}}.
\newblock R package version 0.2.2.

\bibitem[VanderWeele and Ding(2017)]{vanderweele2017sensitivity}
Tyler~J VanderWeele and Peng Ding.
\newblock Sensitivity analysis in observational research: introducing the e-value.
\newblock \emph{Annals of internal medicine}, 167\penalty0 (4):\penalty0 268--274, 2017.
\newblock \doi{10.7326/M16-2607}.

\bibitem[Galili(2015)]{dendextend}
Tal Galili.
\newblock dendextend: an r package for visualizing, adjusting, and comparing trees of hierarchical clustering.
\newblock \emph{Bioinformatics}, 2015.
\newblock \doi{10.1093/bioinformatics/btv428}.
\newblock URL \url{https://doi.org/10.1093/bioinformatics/btv428}.

\end{thebibliography}


\clearpage

\appendix

\section{Supplementary Tables}\label{suptab}
\FloatBarrier
\setcounter{table}{0}
\renewcommand{\thetable}{A\arabic{table}}
\makeatletter
\renewcommand{\theHtable}{A\arabic{table}}
\makeatother

\begin{table}[!htb]
    \centering
    \caption{Estimates for TE, NDE and NIE for each phthalate exposure using the single exposure mediation analysis approach, without including co-exposures as covariates in the mediator and outcome models. The estimates with P-value$\leq0.2$ are highlighted in red. All effects were estimated using the regression-based approach, implemented in the R package \texttt{CMAverse}~\cite{shi2021cmaverse}. Confidence intervals are based on the delta method, and all values are rounded to two decimal places.} 
    \resizebox{0.9\linewidth}{!}{
    \begin{tabular}{ccccccc}
    \hline
    \multirow{2}{*}{\textbf{Exposure}} 
    & \multicolumn{2}{c}{\textbf{TE}} 
    & \multicolumn{2}{c}{\textbf{NDE}} 
    & \multicolumn{2}{c}{\textbf{NIE}} \\
    \cline{2-7}
    & \textbf{Estimate [95\% CI]} & \textbf{P-value (FDR)} 
    & \textbf{Estimate [95\% CI]} & \textbf{P-value (FDR)} 
    & \textbf{Estimate [95\% CI]} & \textbf{P-value (FDR)} \\
    \hline
    MBP
    & {-0.08 [-0.29, 0.13]} & {0.45 (0.61)} 
    & {-0.07 [-0.28, 0.13]} & {0.48 (0.63)} 
    & {-0.01 [-0.03, 0.02]} & {0.61 (0.93)} \\
    MBzP
    & {-0.07 [-0.29, 0.14]} & {0.50 (0.61)} 
    & -0.07 [-0.28, 0.14] & 0.51 (0.63) 
    & {0.00 [-0.03, 0.02]} & {0.85 (0.93)} \\
    MCNP
    & {0.01 [-0.19, 0.22]} & {0.89 (0.89)} 
    & {0.04 [-0.16, 0.25]} & {0.69 (0.76)} 
    & {-0.03 [-0.07, 0.02]} & {0.21 (0.93)} \\
    MCOP
    & {-0.04 [-0.25, 0.16]} & {0.68 (0.74)} 
    & {-0.03 [-0.23, 0.18]} & {0.81 (0.81)} 
    & {-0.02 [-0.05, 0.02]} & {0.29 (0.93)} \\
    MCPP
    & \textcolor{red}{-0.16 [-0.37, 0.05]} & \textcolor{red}{0.14 (0.37)} 
    & \textcolor{red}{-0.16 [-0.37, 0.05]} & \textcolor{red}{0.13 (0.37)} 
    & {0.00 [-0.02, 0.02]} & {0.97 (0.97)} \\
    MECPP
    & {-0.13 [-0.34, 0.08]} & {0.21 (0.37)} 
    & {-0.13 [-0.33, 0.08]} & {0.23 (0.37)} 
    & {0.00 [-0.03, 0.02]} & {0.71 (0.93)} \\
    MEHHP
    & {-0.13 [-0.33, 0.08]} & {0.23 (0.37)} 
    & {-0.12 [-0.33, 0.08]} & {0.23 (0.37)} 
    & {0.00 [-0.03, 0.02]} & {0.85 (0.93)} \\
    MEHP
    & \textcolor{red}{-0.14 [-0.35, 0.07]} & \textcolor{red}{0.18 (0.37)} 
    & {-0.13 [-0.34, 0.07]} & {0.21 (0.37)} 
    & {-0.01 [-0.03, 0.02]} & {0.53 (0.93)} \\
    MEOHP
    & \textcolor{red}{-0.14 [-0.34, 0.07]} & \textcolor{red}{0.19 (0.37)} 
    & \textcolor{red}{-0.13 [-0.34, 0.07]} & \textcolor{red}{0.20 (0.37)} 
    & {-0.01 [-0.03, 0.02]} & {0.67 (0.93)} \\
    MEP
    & {-0.12 [-0.33, 0.08]} & {0.24 (0.37)} 
    & \textcolor{red}{-0.15 [-0.36, 0.05]} & \textcolor{red}{0.14 (0.37)} 
    & \textcolor{red}{0.03 [-0.01, 0.08]} & \textcolor{red}{0.16 (0.93)} \\
    MiBP
    & \textcolor{red}{-0.18 [-0.39, 0.02]} & \textcolor{red}{0.08 (0.37)} 
    & \textcolor{red}{-0.19 [-0.40, 0.01]} & \textcolor{red}{0.06 (0.37)} 
    & {0.01 [-0.02, 0.04]} & {0.49 (0.93)} \\
    \hline
    \end{tabular}}
    \label{tab:data_ind}
\end{table}

\begin{table}[!htb]
    \centering
    \caption{Estimates for TE, NDE and NIE for each phthalate exposure using the single exposure mediation analysis approach, with co-exposures included as covariates in the mediator and outcome models. The estimates with P-value$\leq0.2$ are highlighted in red. All effects were estimated using the regression-based approach, implemented in the R package \texttt{CMAverse}~\cite{shi2021cmaverse}. Confidence intervals are based on the delta method, and all values are rounded to two decimal places.} 
    \resizebox{0.9\linewidth}{!}{
    \begin{tabular}{ccccccc}
    \hline
    \multirow{2}{*}{\textbf{Exposure}} 
    & \multicolumn{2}{c}{\textbf{TE}} 
    & \multicolumn{2}{c}{\textbf{NDE}} 
    & \multicolumn{2}{c}{\textbf{NIE}} \\
    \cline{2-7}
    & \textbf{Estimate [95\% CI]} & \textbf{P-value (FDR)} 
    & \textbf{Estimate [95\% CI]} & \textbf{P-value (FDR)} 
    & \textbf{Estimate [95\% CI]} & \textbf{P-value (FDR)} \\
    \hline
    MBP
    & {0.13 [-0.17, 0.43]} & {0.40 (0.89)} 
    & {0.17 [-0.13, 0.46]} & {0.27 (0.75)} 
    & {-0.04 [-0.10, 0.02]} & {0.22 (0.49)} \\
    MBzP
    & {-0.02 [-0.27, 0.23]} & {0.90 (0.97)} 
    & -0.02 [-0.26, 0.23] & 0.89 (0.95) 
    & {0.00 [-0.04, 0.04]} & {0.96 (0.96)} \\
    MCNP
    & {0.15 [-0.12, 0.42]} & {0.27 (0.89)} 
    & \textcolor{red}{0.19 [-0.07, 0.46]} & \textcolor{red}{0.15 (0.56)} 
    & \textcolor{red}{-0.04 [-0.10, 0.02]} & \textcolor{red}{0.16 (0.49)} \\
    MCOP
    & {0.08 [-0.24, 0.39]} & {0.64 (0.89)} 
    & {0.11 [-0.21, 0.42]} & {0.50 (0.92)} 
    & {-0.03 [-0.09, 0.03]} & {0.31 (0.49)} \\
    MCPP
    & \textcolor{red}{-0.24 [-0.54, 0.06]} & \textcolor{red}{0.12 (0.89)} 
    & \textcolor{red}{-0.28 [-0.59, 0.02]} & \textcolor{red}{0.06 (0.55)} 
    & \textcolor{red}{0.04 [-0.02, 0.11]} & \textcolor{red}{0.19 (0.49)} \\
    MECPP
    & {-0.01 [-0.59, 0.56]} & {0.97 (0.97)} 
    & {-0.02 [-0.59, 0.55]} & {0.95 (0.95)} 
    & {0.01 [-0.09, 0.10]} & {0.87 (0.96)} \\
    MEHHP
    & {0.16 [-0.73, 1.04]} & {0.73 (0.89)} 
    & {0.07 [-0.81, 0.94]} & {0.88 (0.95)} 
    & {0.09 [-0.08, 0.26]} & {0.31 (0.49)} \\
    MEHP
    & {-0.08 [-0.42, 0.25]} & {0.63 (0.89)} 
    & {-0.06 [-0.40, 0.27]} & {0.71 (0.95)} 
    & {-0.02 [-0.08, 0.04]} & {0.54 (0.66)} \\
    MEOHP
    & {-0.19 [-1.07, 0.69]} & {0.67 (0.89)} 
    & {-0.12 [-0.99, 0.75]} & {0.79 (0.95)} 
    & {-0.07 [-0.23, 0.09]} & {0.38 (0.52)} \\
    MEP
    & {-0.07 [-0.28, 0.15]} & {0.53 (0.89)} 
    & {-0.10 [-0.31, 0.11]} & {0.36 (0.78)} 
    & \textcolor{red}{0.03 [-0.01, 0.08]} & \textcolor{red}{0.17 (0.49)} \\
    MiBP
    & \textcolor{red}{-0.20 [-0.48, 0.08]} & \textcolor{red}{0.16 (0.89)} 
    & \textcolor{red}{-0.23 [-0.51, 0.04]} & \textcolor{red}{0.10 (0.55)} 
    & {0.03 [-0.02, 0.09]} & {0.24 (0.49)} \\
    \hline
    \end{tabular}}
    \label{tab:data_indco}
\end{table}

\begin{table}[!htb]
    \centering
    \caption{Estimates for TE, NDE and NIE for the first five PCs derived from the 11 phthalate exposures in the PROTECT dataset. The estimates with P-value$\leq0.2$ are highlighted in red. All effects were estimated using the regression-based approach, implemented in the R package \texttt{CMAverse}~\cite{shi2021cmaverse}. Confidence intervals are based on the delta method, and all values are rounded to two decimal places.} 
    \resizebox{0.9\linewidth}{!}{
    \begin{tabular}{ccccccc}
    \hline
    \multirow{2}{*}{\textbf{Principal Components}} 
    & \multicolumn{2}{c}{\textbf{TE}} 
    & \multicolumn{2}{c}{\textbf{NDE}} 
    & \multicolumn{2}{c}{\textbf{NIE}} \\
    \cline{2-7}
    & \textbf{Estimate [95\% CI]} & \textbf{P-value (FDR)} 
    & \textbf{Estimate [95\% CI]} & \textbf{P-value (FDR)} 
    & \textbf{Estimate [95\% CI]} & \textbf{P-value (FDR)} \\
    \hline
    PC1
    & \textcolor{red}{0.07 [-0.02, 0.16]} & \textcolor{red}{0.14 (0.69)} 
    & \textcolor{red}{0.07 [-0.03, 0.16]} & \textcolor{red}{0.16 (0.57)} 
    & {0.00 [-0.01, 0.02]} & {0.59 (0.67)} \\
    PC2
    & {-0.05 [-0.21, 0.12]} & {0.56 (0.80)} 
    & {-0.07 [-0.24, 0.09]} & {0.40 (0.66)}
    & \textcolor{red}{0.02 [-0.01, 0.06]} & \textcolor{red}{0.19 (0.52)} \\
    PC3
    & {0.10 [-0.10, 0.30]} & {0.33 (0.80)} 
    & {0.12 [-0.08, 0.32]} & {0.23 (0.57)} 
    & {-0.03 [-0.06, 0.01]} & {0.21 (0.52)} \\
    PC4
    & {0.05 [-0.17, 0.28]} & {0.64 (0.80)} 
    & {0.07 [-0.15, 0.29]} & {0.55 (0.69)} 
    & {-0.01 [-0.05, 0.02]} & {0.43 (0.67)} \\
    PC5
    & {-0.01 [-0.27, 0.24]} & {0.91 (0.91)} 
    & {-0.01 [-0.26, 0.24]} & {0.95 (0.95)} 
    & {-0.01 [-0.04, 0.03]} & {0.67 (0.67)} \\
    \hline
    \end{tabular}}s
    \label{tab:pca_data}
\end{table}

\begin{table}[!htb]
    \centering   
    \caption{Estimated TE, NDE and NIE using the ERS in the PROTECT dataset. Effects represent the change in head circumference associated with an increase in the ERS from the 25th to the 75th percentile in the analysis set. Two ERS specifications are shown: one including only main effects, and another including both main effects and higher-order terms (squared and interaction terms). All effects are estimated using the regression-based approach, implemented in the R package \texttt{CMAverse}~\cite{shi2021cmaverse}. Confidence intervals are based on the delta method. All values are rounded to two decimal places.}
    \begin{tabular}{ccccc}
    \toprule
        \multicolumn{5}{c}{\textbf{Main Effects Only}} \\
    \midrule
        \textbf{Effects} & \textbf{Estimates} & \textbf{SE} & \textbf{95\% CI} & \textbf{P-value}\\  
    \midrule
        TE & 0.14 & 0.18 & (-0.21, 0.49) & 0.44\\
        NDE & 0.12 & 0.18 & (-0.23, 0.48) & 0.50\\
        NIE & 0.02 & 0.03 & (-0.04, 0.07) & 0.56\\
    \midrule
        \multicolumn{5}{c}{\textbf{Main Effects, Squared Terms and Interactions}} \\
    \midrule
        \textbf{Effects} & \textbf{Estimates} & \textbf{SE} & \textbf{95\% CI} & \textbf{P-value}\\     
    \midrule
        TE & 0.17 & 0.18 & (-0.19, 0.52) & 0.36\\
        NDE & 0.17 & 0.18 & (-0.19, 0.52) & 0.36\\
        NIE & 0.00 & 0.02 & (-0.03, 0.03) & 0.94\\
    \bottomrule
    \end{tabular}
    \label{tab:ers_data}
\end{table}

\begin{table}[!htb]
    \centering   
    \caption{Estimated global NIEs of the phthalate mixture on head circumference Z-score in the PROTECT dataset. All estimates and intervals are rounded to two decimal places.}
    \begin{tabular}{ccc}
    \hline
    Method & Estimate & 95\% CI/CrI \\
    \hline
    SE-MA (w/ co-exposures) & 0.00 & (-0.30, 0.30)\\
    PC-MA & -0.02 & (-0.09, 0.05)\\
    ERS-MA (main effect only) & 0.02 & (-0.04, 0.13)\\
    ERS-MA (main effect w/ high-order terms) & 0.00 & (-0.05, 0.06)\\
    BKMR-CMA (Component-wise) & 0.00 & (-0.44, 0.45) \\
    BKMR-CMA (Hierarchical) & -0.02 & (-0.40, 0.39)\\
    \hline
    \end{tabular}
    \label{tab:nie_data}
\end{table}

\FloatBarrier 

\clearpage

\section{Supplementary Figures}

\setcounter{figure}{0}
\renewcommand{\thefigure}{B\arabic{figure}}
\makeatletter
\renewcommand{\theHfigure}{B\arabic{figure}}
\makeatother

\begin{figure}[!htbp]
    \centering
    \begin{tikzpicture}[
        node distance = 2cm and 2.5cm,
        every node/.style = {align=center},
        arrow/.style = {-{Stealth}, thick}
    ]    
    \node[draw=brown, rounded corners] (X) at (0,0) {PC Scores \\ 
    $\bS=f(\X)_{(n\times l)}$};
    \node[draw=brown, rounded corners] (M) [right=of X] {Mediator \\ 
    $M_{(n\times1)}$};
    \node[draw=brown, rounded corners] (Y) [right=of M] {Outcome \\ 
    $Y_{(n\times1)}$};
    \node[draw=brown, rounded corners] (C) [above=of M, yshift=1cm] {Confounders \\
    $\C_{(n\times s)}$};
    
    \draw [arrow] (X) -- (M);
    \draw [arrow] (M) -- (Y);
    \draw [arrow] (X) to[bend right] (Y);
    \draw [arrow] (C) -- (X);
    \draw [arrow] (C) -- (M);
    \draw [arrow] (C) -- (Y);
    
    \end{tikzpicture}
    \caption{Directed acyclic graph illustrating the use of principal component analysis in mediation analysis of exposure mixtures. The original exposures $\X$ are transformed into principal component (PC) scores $\bS=f(\X)$, which serve as dimension-reduced representations of the exposure mixture in the mediator and outcome models. The relationships among PC scores, mediator, outcome, and confounders are modeled without directly including the original exposures.}
    \label{fig:pca_dag}
\end{figure}
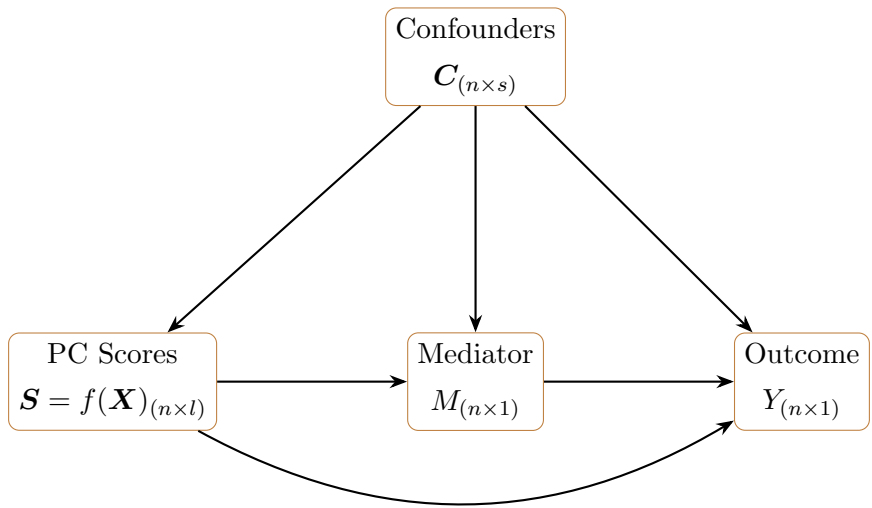

\begin{figure}[!htb]
    \centering
    \includegraphics[width=0.9\linewidth]{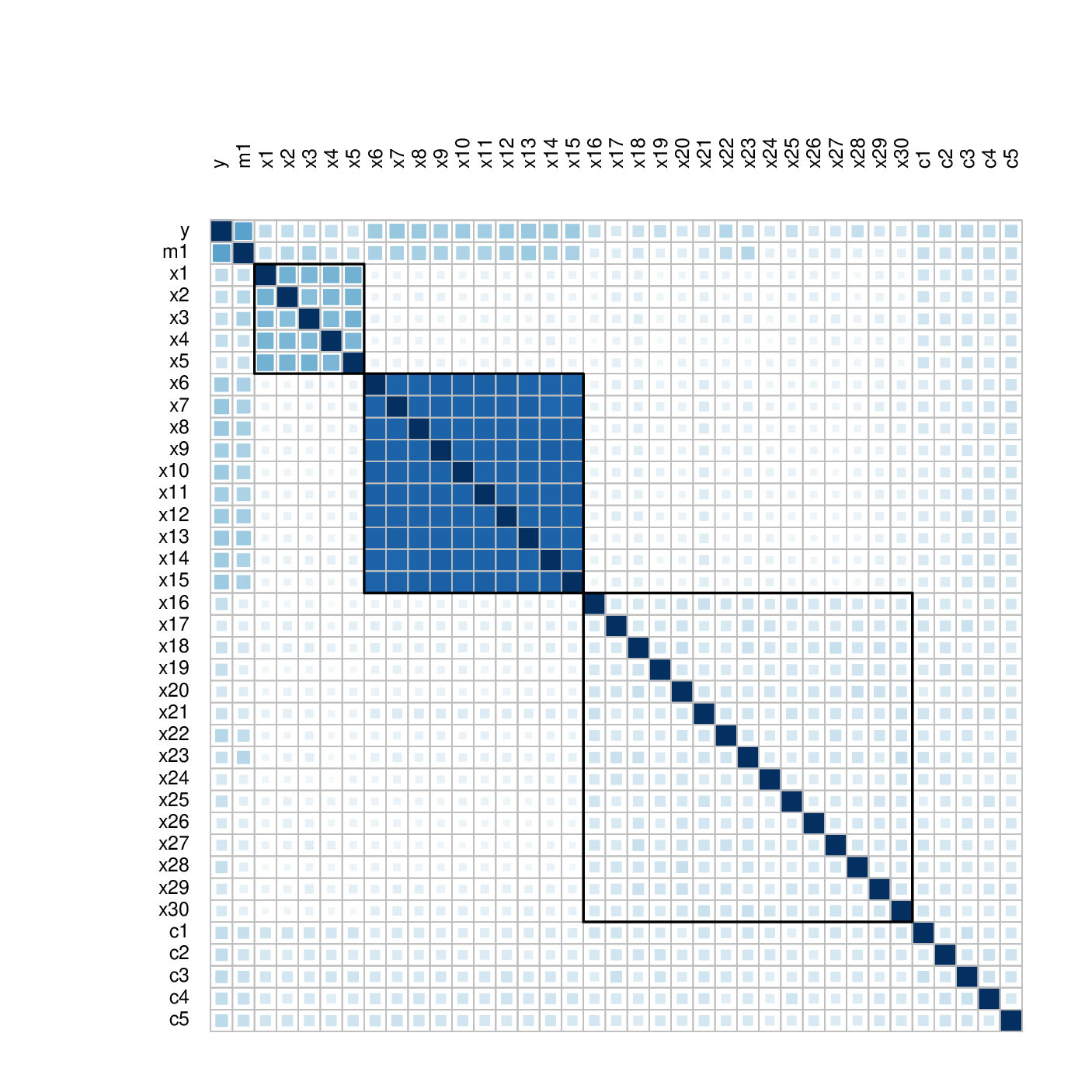}
    \caption{Pearson correlation matrix from one simulated dataset under the scenario with $n=2,500$ and $R^2_M=0.4$. The exposures are grouped into three distinct mixture components with varying within-group correlations. The confounders also show weak correlations among themselves. The intercept term is excluded from this plot, as it is constant across observations with a value of 1. Darker cells represent stronger correlations, while lighter cells represent weaker or no correlations. The black squares highlight the three exposure mixture components.}
    \label{fig:corplot}
\end{figure}

\begin{figure}[!htb]
    \centering
    \includegraphics[width=0.9\linewidth]{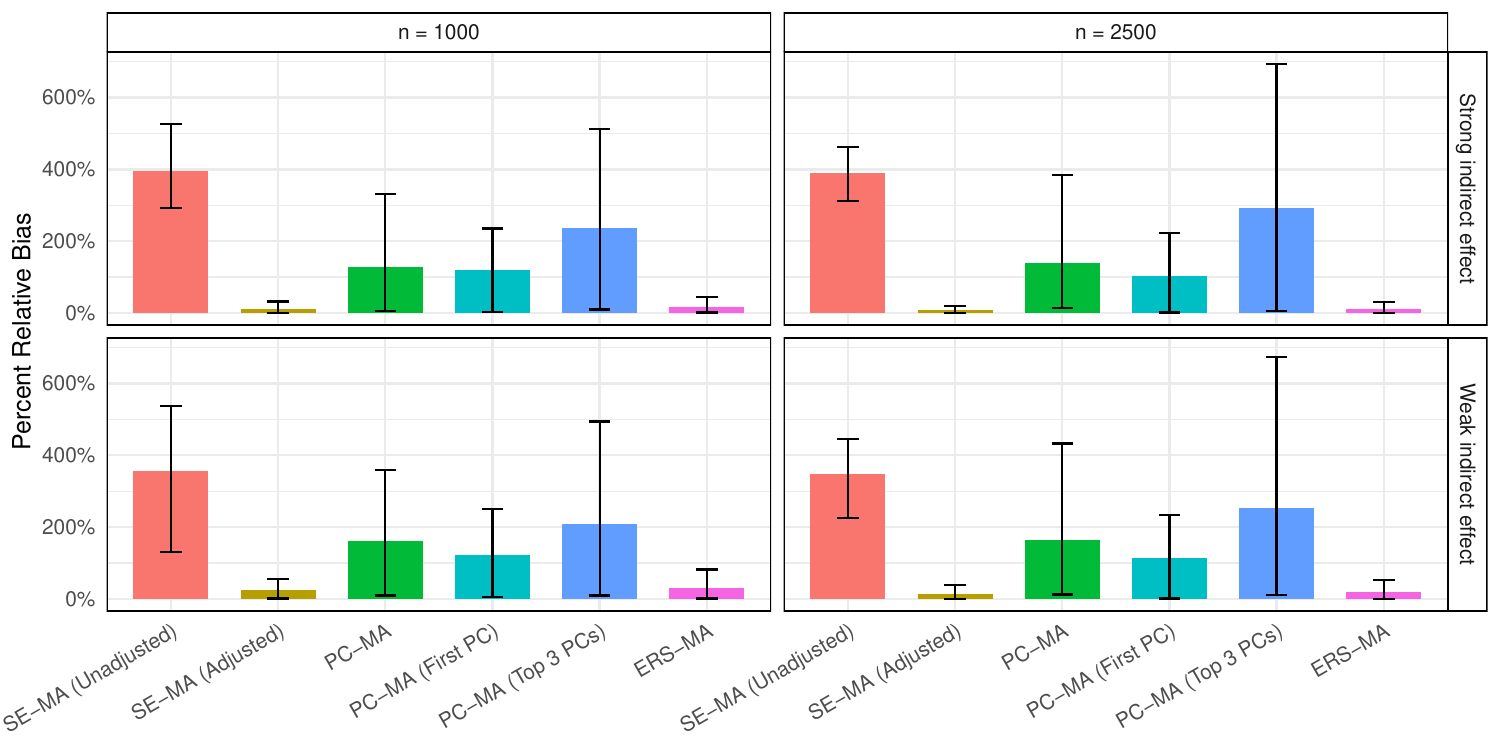}
    \caption{Percent relative bias in estimating the global natural indirect effect (NIE) across simulation scenarios. Relative bias is computed as the absolute value of the average deviation between the estimated and the true global NIE and expressed as a percentage. Methods compared include unadjusted and adjusted SE-MA, PC-MA (using all PCs, the first PC, or top 3 PCs), and ERS-MA with ERS constructed from main effects only.}
    \label{fig:sim_rbias}
\end{figure}

\begin{figure}[!htb]
    \centering
    \includegraphics[width=0.9\linewidth]{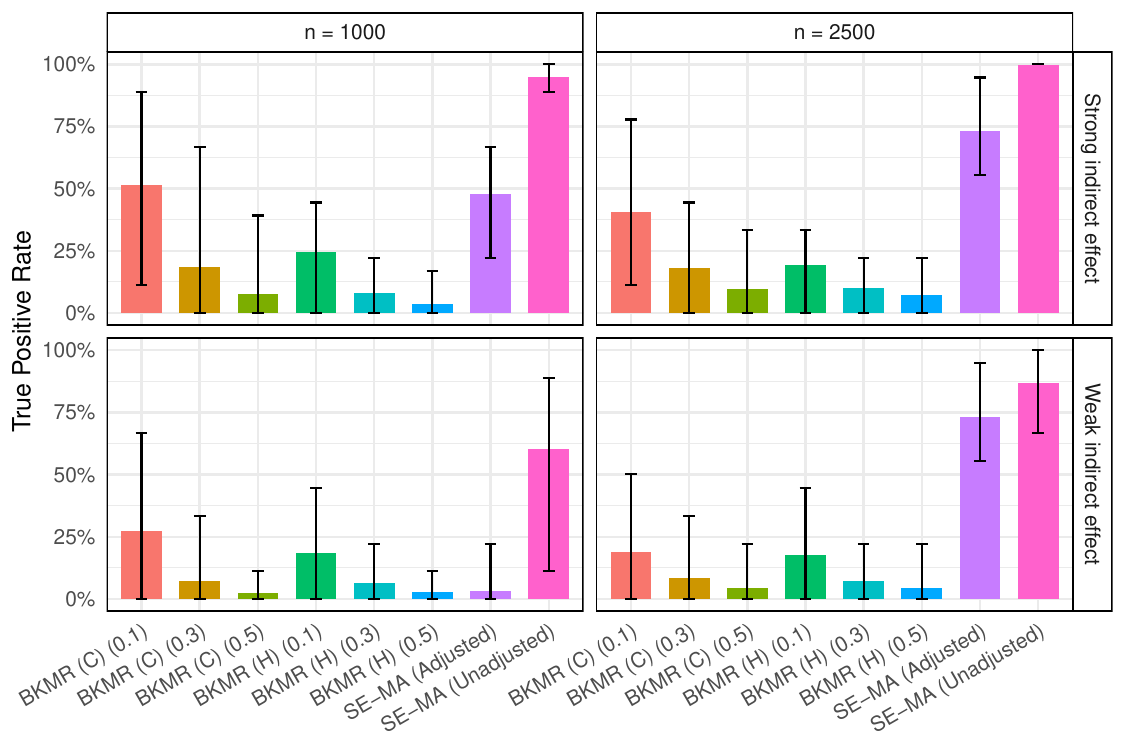}
    \caption{True positive rates (TPRs) for identifying exposures with non-zero natural indirect effects (NIEs) across simulation scenarios. Each panel corresponds to a combination of sample size ($n=1,000$ or $2,500$) and mediator model strength ($R^2_M=0.1$ or $0.4$). Bars show average TPRs across 100 simulated datasets, with error bars representing standard deviations. For BKMR-CMA, we consider three posterior inclusion probability (PIP) cutoffs ($0.1$, $0.3$, $0.5$) applied to both component-wise (C) and hierarchical (H) models. For SE-MA (with and without adjusting for co-exposures), exposures are flagged when the FDR for the NIE is less than 0.05. Higher TPR values reflect greater sensitivity in detecting exposures with non-zero NIEs.}
    \label{fig:sim_tpr}
\end{figure}

\begin{figure}[!htb]
    \centering
    \includegraphics[width=0.9\linewidth]{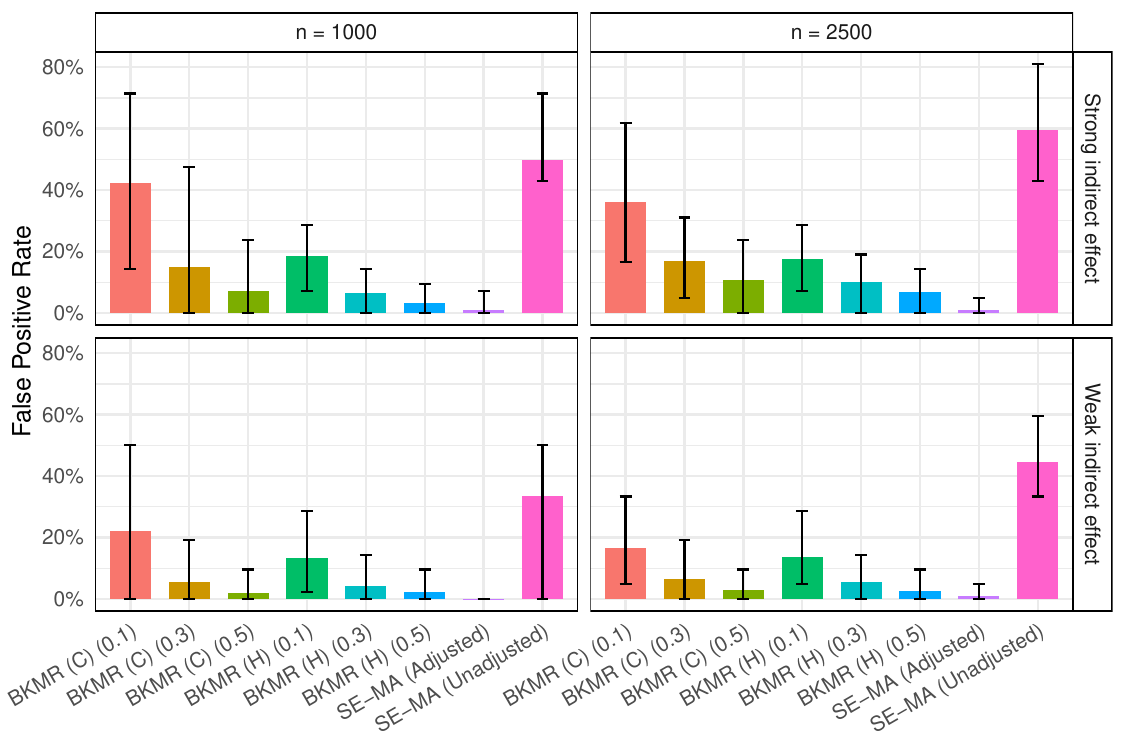}
    \caption{False positive rates (FPRs) for identifying exposures with null natural indirect effects across simulation scenarios. Simulation settings match those in Figure~\ref{fig:sim_tpr}. FPR is defined as the proportion of truly null exposures incorrectly identified as having non-zero NIEs.}
    \label{fig:sim_fpr}
\end{figure}

\begin{figure}[!htbp]
    \centering
    \includegraphics[width=0.7\linewidth]{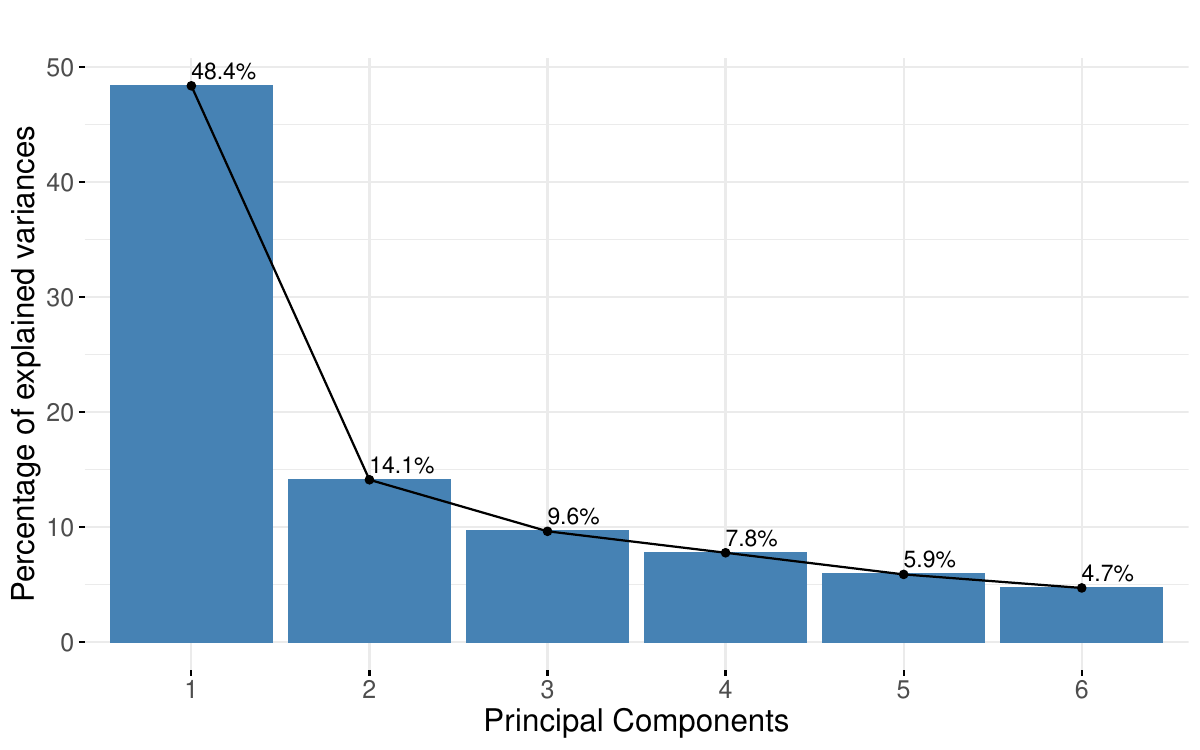}
    \caption{Scree plot of the principal components for the 11 phthalate exposures in the PROTECT dataset. The proportion of variance explained is labeled above each principal component. The plot is generated using the R package \texttt{factoextra}~\cite{factoextra}.}
    \label{fig:pca_data_scree}
\end{figure}

\begin{figure}[!htbp]
    \centering
    \includegraphics[width=0.7\linewidth]{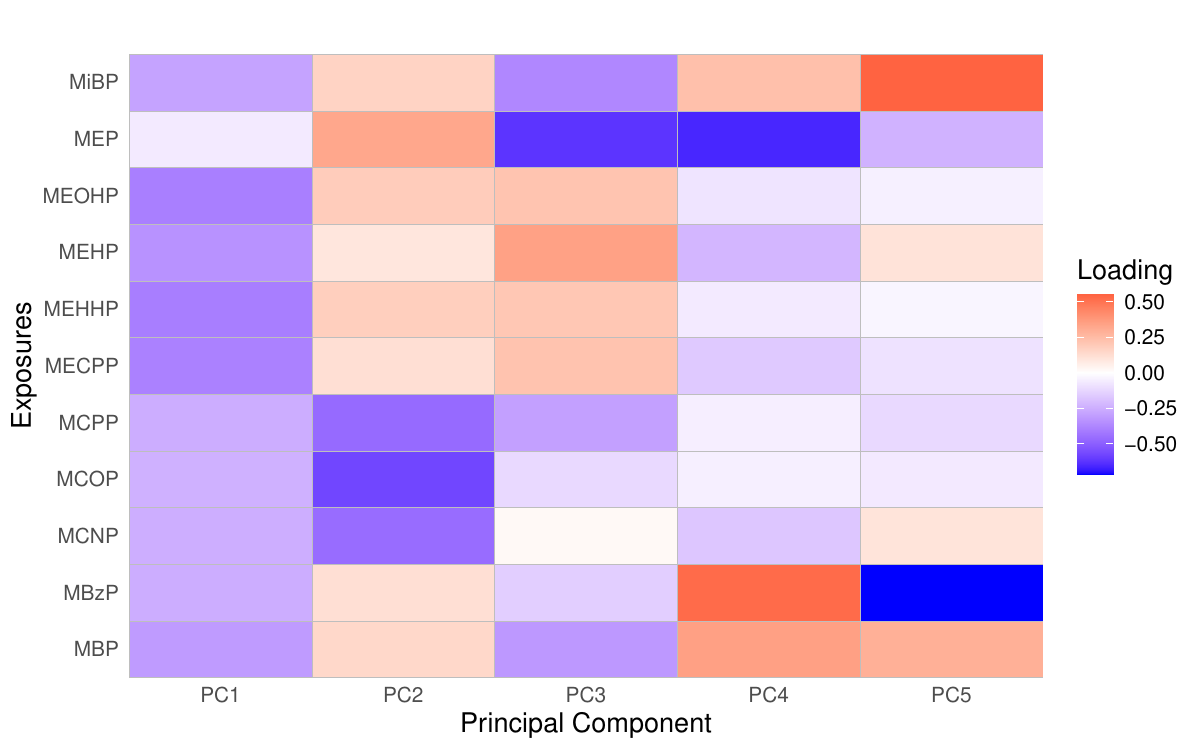}
    \caption{Heatmap of the principal component loadings calculated from the 11 phthalate exposures of the PROTECT dataset. The first 5 principal components that are included in subsequent analysis are presented in the heatmap.}
    \label{fig:pca_data_heat}
\end{figure}

\begin{figure}[!htbp]
    \centering
    \includegraphics[width=0.8\linewidth]{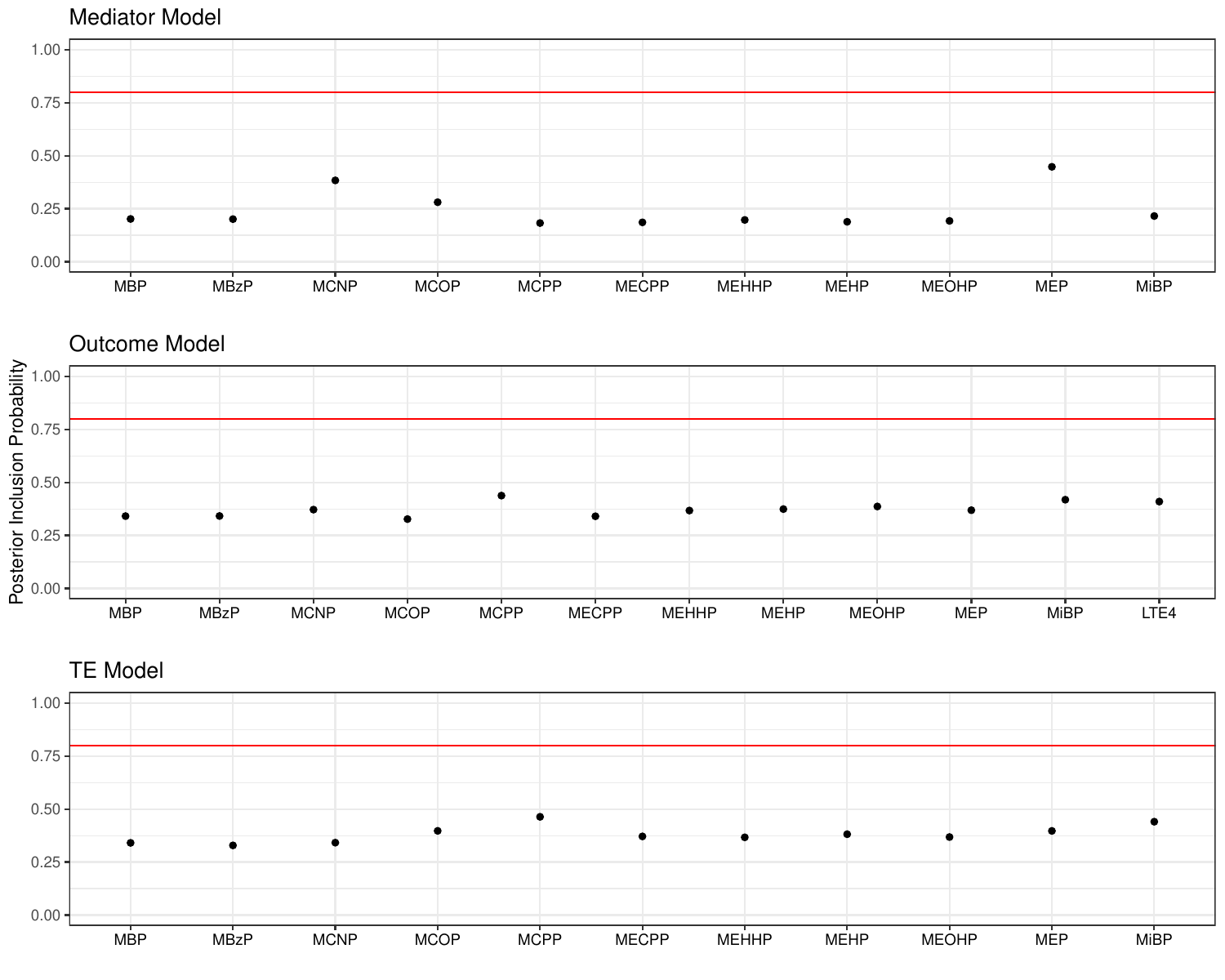}
    \caption{Posterior inclusion probabilities (PIPs) of individual exposures in the three BKMR models. A reference line at $\text{PIP} = 0.8$ is provided to indicate a commonly used threshold for strong inclusion probability.}
    \label{fig:bkmr_data_pip}
\end{figure}

\begin{figure}[!htbp]
    \centering
    \includegraphics[width=0.7\linewidth]{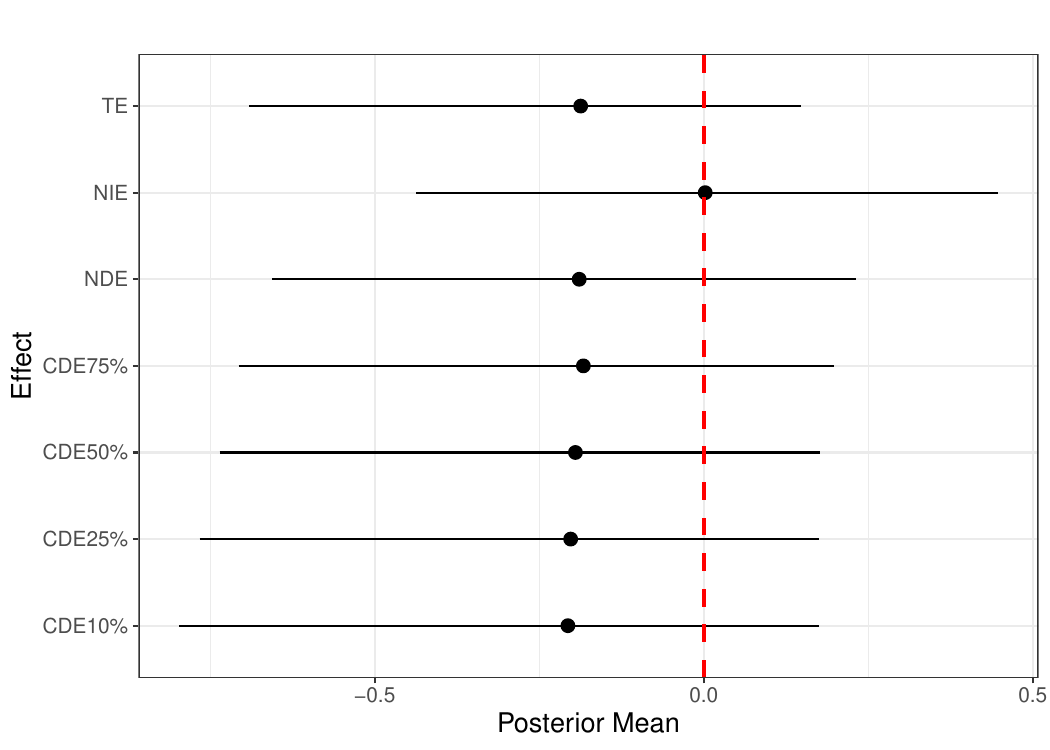}
    \caption{Forest plot of causal mediation effect estimates from BKMR-CMA applied to the PROTECT dataset, using component-wise variable selection. Each effect is estimated under the setting where all exposures jointly change from the 25th to the 75th percentile, with confounders fixed at their mean values. The plot displays the posterior mean and 95\% credible interval for each effect, with a reference line at zero to aid interpretation. CDE10\%, CDE25\%, CDE50\%, and CDE75\% denote controlled direct effects evaluated at the 10th, 25th, 50th, and 75th percentiles of the mediator, respectively.}
    \label{fig:bkmr_data_est}
\end{figure}

\begin{figure}[!htbp]
    \centering
    \includegraphics[width=0.7\linewidth]{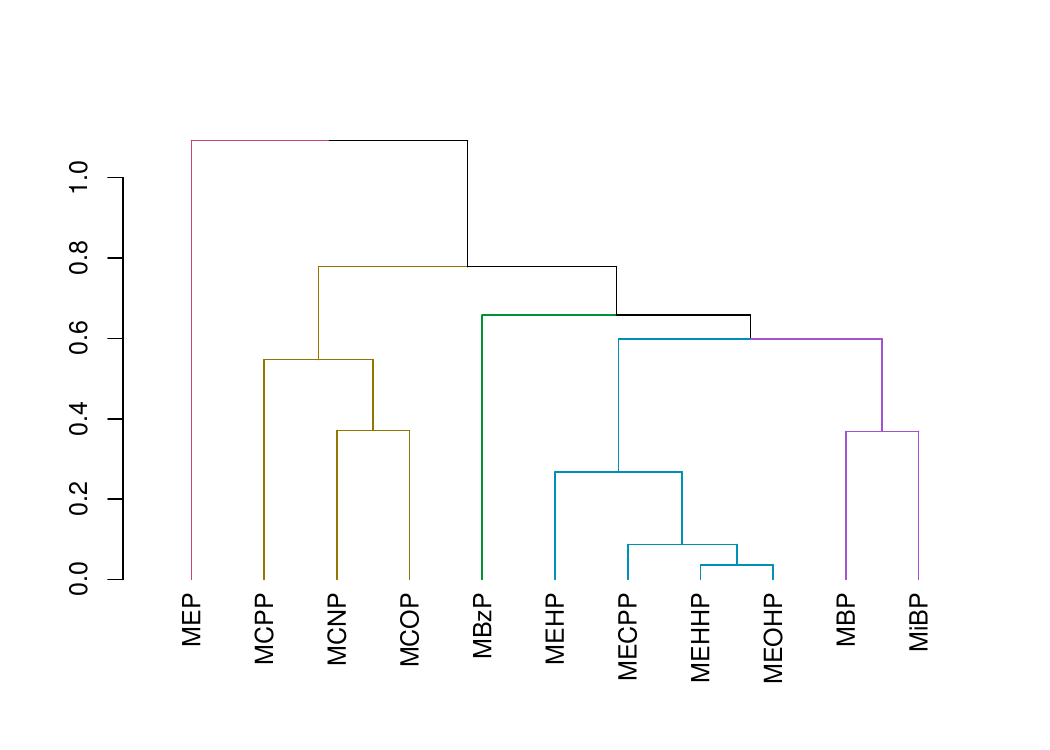}
    \caption{Clustering dendrogram of the hierarchical clustering results of the phthalate exposures. The hierarchical clustering process is performed with the complete linkage and correlation matrix of the exposures as the measurement for similarity between exposures. Exposures of different clustering groups are highlighted in different colors. The left bar indicates the calculated similarity between exposures. The dendrogram is plotted with the R package \texttt{dendextend}.~\cite{dendextend}}
    \label{fig:bkmr_data_clust}
\end{figure}

\begin{figure}[!htbp]
    \centering
    \includegraphics[width=0.8\linewidth]{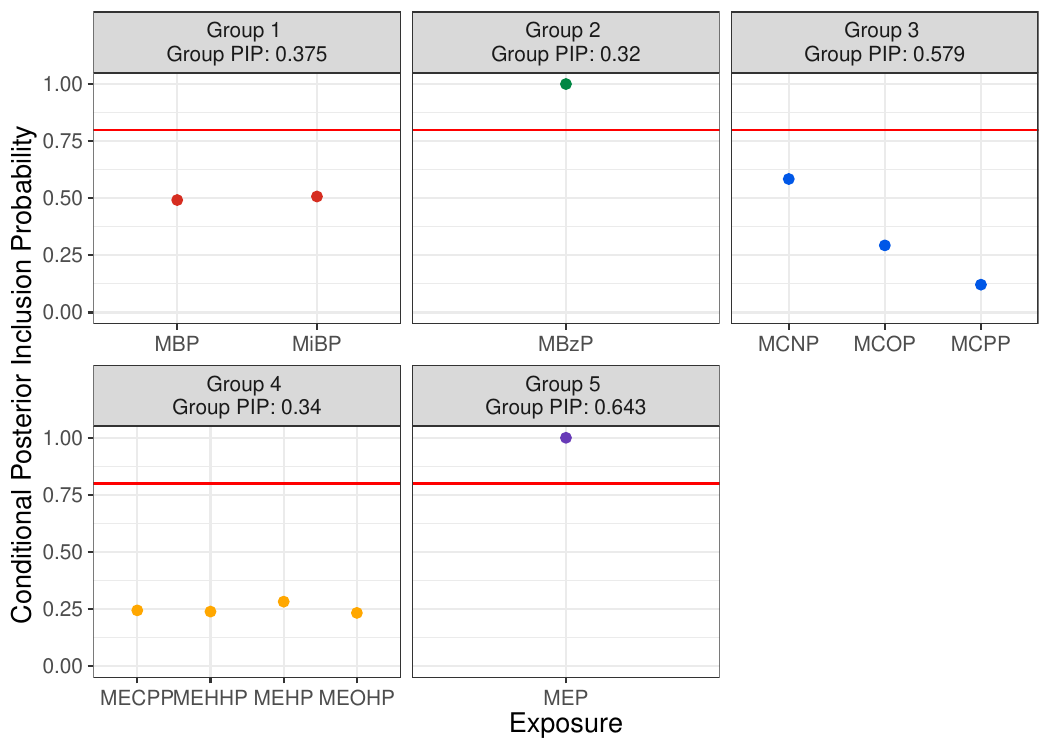}
    \caption{Posterior inclusion probabilities (PIPs) of the BKMR mediator model with the hierarchical variable selection method for the PROTECT dataset. Group PIP of different exposure clustering groups are presented in the header of each facet. Exposures of different clustering groups are highlighted in different colors. Reference lines are drawn on the conditional $\text{PIP} = 0.8$.}
    \label{fig:bkmr_data_hiermedpip}
\end{figure}

\begin{figure}[!htbp]
    \centering
    \includegraphics[width=0.8\linewidth]{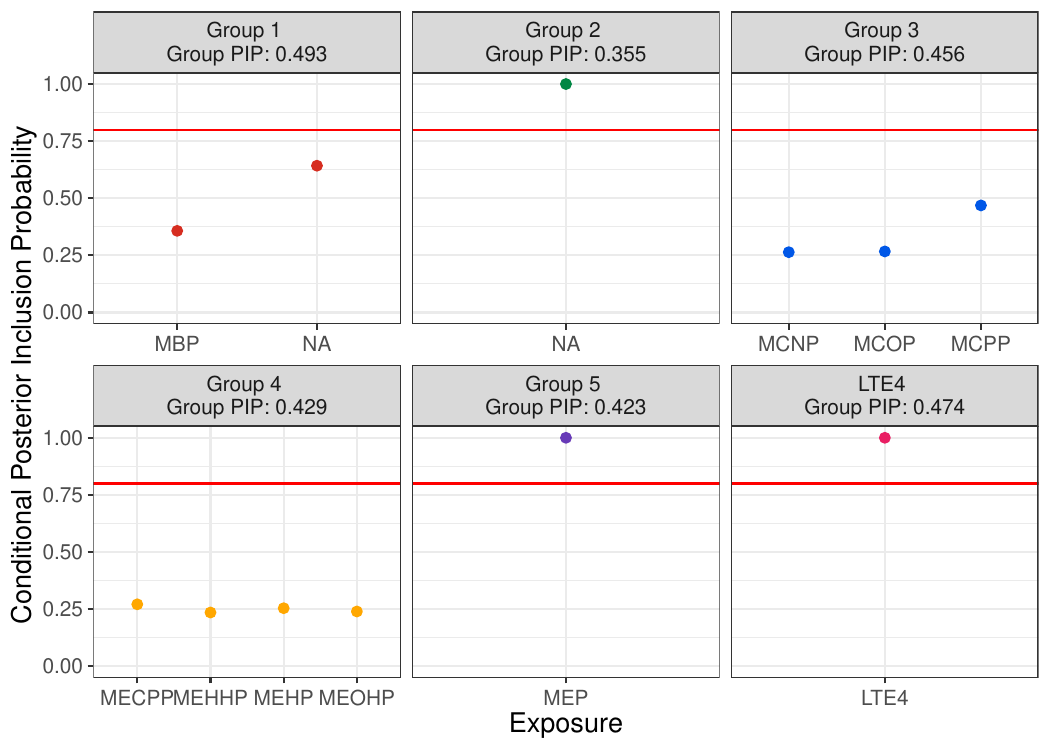}
    \caption{Posterior inclusion probabilities (PIPs) of the BKMR outcome model with the hierarchical variable selection method for the PROTECT dataset. Group PIP of different exposure clustering groups are presented in the header of each facet. Exposures of different clustering groups are highlighted in different colors. Reference lines are drawn on the conditional $\text{PIP} = 0.8$.}
    \label{fig:bkmr_data_hieroutpip}
\end{figure}

\begin{figure}[!htbp]
    \centering
    \includegraphics[width=0.7\linewidth]{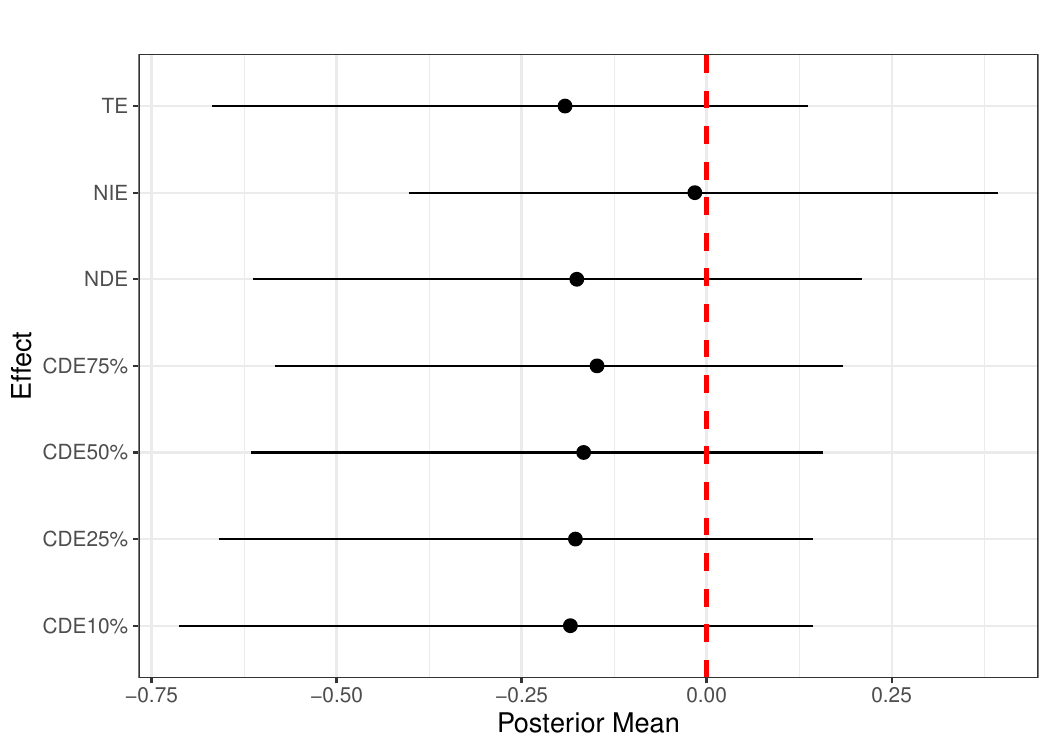}
    \caption{Forest plot of causal mediation effect estimates from BKMR-CMA applied to the PROTECT dataset, using hierarchical variable selection. Each effect is estimated under the setting where all exposures jointly change from the 25th to the 75th percentile, with confounders fixed at their mean values. The plot displays the posterior mean and 95\% credible interval for each effect, with a reference line at zero to aid interpretation. CDE10\%, CDE25\%, CDE50\%, and CDE75\% denote controlled direct effects evaluated at the 10th, 25th, 50th, and 75th percentiles of the mediator, respectively.}
    \label{fig:bkmr_data_hier_est}
\end{figure}

\end{document}